\documentclass[journal=asbcd6,manuscript=article]{achemso}
\setkeys{acs}{keywords = true}
\usepackage{float}
\usepackage{graphicx}
\usepackage{color}
\usepackage{colortbl}
\usepackage{subfigure}
\usepackage{amsmath}
\usepackage{algorithmic}
\usepackage{algorithm}
\usepackage{flushend}
\usepackage{amssymb}
\usepackage{multirow}
\usepackage{CJK}
\usepackage{array}
\usepackage{amsmath,bm}
\usepackage{makecell}

\newtheorem{Def}{Definition}

\newtheorem{Rem}{Remark}
\newtheorem{proof}{Proof}
\usepackage[version=3]{mhchem} 
\usepackage[T1]{fontenc}       

\author{Chuan Zhang}
\email{chzhang@seu.edu.cn}
\affiliation[LEADS]
{Lab of Efficient Architectures for Digital-communication and Signal-processing (LEADS)}
\alsoaffiliation[Quantum]
{Quantum Information Center of Southeast University}
\alsoaffiliation[NCRL]
{National Mobile Communications Research Laboratory, Southeast University, China}
\author{Ziyuan Shen}
\affiliation[LEADS]
{Lab of Efficient Architectures for Digital-communication and Signal-processing (LEADS)}
\alsoaffiliation[Quantum]
{Quantum Information Center of Southeast University}
\alsoaffiliation[NCRL]
{National Mobile Communications Research Laboratory, Southeast University, China}
\author{Wei Wei}
\affiliation[Chemistry]
{State Key Laboratory of Coordination Chemistry, School of Chemistry and Chemical Engineering, Nanjing University, China}
\alsoaffiliation[Biology]
{State Key Laboratory of Pharmaceutical Biotechnology, School of Life Sciences, Nanjing
University, China}
\author{Jing Zhao}
\affiliation[Chemistry]
{State Key Laboratory of Coordination Chemistry, School of Chemistry and Chemical Engineering, Nanjing University, China}
\alsoaffiliation[Biology]
{State Key Laboratory of Pharmaceutical Biotechnology, School of Life Sciences, Nanjing
University, China}
\author{Zaichen Zhang}
\affiliation[Quantum]
{Quantum Information Center of Southeast University}
\alsoaffiliation[NCRL]
{National Mobile Communications Research Laboratory, Southeast University, China}
\author{Xiaohu You}
\affiliation[NCRL]
{National Mobile Communications Research Laboratory, Southeast University, China}

\title[An \textsf{achemso} demo]
  {Molecular Computing for Markov Chains\footnote{Chuan Zhang and Ziyuan Shen contributed equally to this work.}}

\keywords{molecular computing; DNA strand displacement; Markov chain; mass action kinetics; Gillespie algorithm}
\begin{document}
\begin{abstract}
In this paper, it is presented a methodology for implementing arbitrarily constructed time-homogenous Markov chains with biochemical systems. Not only discrete but also continuous-time Markov chains are allowed to be computed. By employing chemical reaction networks (CRNs) as a programmable language, molecular concentrations serve to denote both input and output values. One reaction network is elaborately designed for each chain. The evolution of species' concentrations over time well matches the transient solutions of the target continuous-time Markov chain, while equilibrium concentrations can indicate the steady state probabilities. Additionally, second-order Markov chains are considered for implementation, with bimolecular reactions rather that unary ones. An original scheme is put forward to compile unimolecular systems to DNA strand displacement reactions for the sake of future physical implementations. Deterministic, stochastic and DNA simulations are provided to enhance correctness, validity and feasibility.
\end{abstract}

\section{Introduction}
By far, the exploitation and application of traditional computing equipment, such as silicon-based devices, has reached its peak. This urges the need of new material for possibly better computation performance or different application scenarios. Capable of exhibiting abundant dynamic behaviors, chemical reaction networks (CRNs) turn out to be a programmable language, prompting molecular scale material to become a highly promising candidate. As a parallel system in nature, CRNs possess the potential to handle large-scale and sophisticated computations. The past few decades have seen a groundswell of interest in molecular computing no matter concerning academy or industry \cite{bennett1982thermodynamics,stemmer1995evolution,puaun2002guide,lund2010molecular}, with scientists trying to reveal the natural programmability of CRNs. A wealth of research is of primary interest in exploring the potential computational power of biological molecules by implementing digital logic, signal processing and functions \cite{chen2014deterministic,jiang2013digital,jiang2011synchronous,kharam2011binary,salehi2014asynchronous,jiang2013discrete,salehi2015molecular,salehi2016chemical}. Some other researchers are inclined to biochemically address computationally intractable and complex problems \cite{adleman1994molecular,ouyang1997dna,salehi2015markov,cardona2005markov}. Even more remarkable works \cite{berry1992chemical,rothemund1995dna,magnasco1997chemical,liekens2007turing,soloveichik2008computation,hjelmfelt1991chemical}, strongly dig out and prove the Turing-universal quality of chemical reaction networks.

Over the past five decades, these works \cite{mcquarrie1967stochastic,van1995stochastic,anderson2011continuous} have been pursuing to building stochastic models for chemical kinetics, among which Markov chains play an important role. In the special field of DNA, Kannan\cite{kannanmarkov} utilizes Markov chains to provide statistical analysis of genome data. While stochastic processes that describe existing chemical systems have been systematically established, the inverse problem of computing stochastic networks by molecular reactions remains unsolved. Only a few people have considered this question in spite of so many extraordinary studies on molecular computing.

Apart from application in chemistry, Markov chains have been successfully applied to a wide range of areas such as digital communications, social networks, finance, and sports. Thus, our work anticipates a main focus on Markov chain related molecular computation. In fact, Cardon \cite{cardona2005markov} and Salehi \cite{salehi2015markov} have already challenged this topic: estimating the steady state distribution of any discrete-time Markov chain (DTMC) by DNA computing. Exactly belonging to the realm of molecular computing, such idea is greatly updated and innovative. In Cardon's paper \cite{cardona2005markov}, DNA strands are used to represent Markov chains' vertexes and edges directly, while in Salehi's \cite{salehi2015markov}, hypothetical reactions are firstly designed. Besides the stationary behavior, the transient behavior---$n$-step transition probabilities of DTMC, is well synthesized\cite{shen2016synthesis}. Unfortunately, none of the aforementioned approaches have made allowance for continuous-time Markov chains (CTMC) or higher-order Markov chains, of which our real life is a closer archetype.

Therefore, this paper attempts to tackle the issue from a more general standpoint. A straightforward and elegant way is proposed for designing CRNs with the functionality of computing not only DTMC but also CTMC and second-order Markov chains. Similar to Salehi\cite{salehi2015markov}, each state is modeled by a unique molecular type. Instead of utilizing control molecules to regulate transitions as in paper\cite{salehi2015markov}, we model state transitions by various rate constants to reduce the number of needed molecular species and for convenience of DNA implementation. Hence, unimolecular reactions are designed for first-order Markov chains and bimolecular reactions serve to compute second-order ones. Different from electronic systems, molecular systems are usually designed with desired results indicated by concentrations as opposed to voltage. And as such, in our methodology, input and output values, which are a Markov chain's initial distribution and steady state probabilities respectively, are both represented by molecular concentrations. Besides, from simulation results, transient solutions of continuous-time Markov chains can be creditably predicted by the evolution of various species' concentrations over time. Both deterministic and stochastic simulations are provided to validate accuracy. Ordinary differential equations (ODEs) analysis is given for CTMC to prove infallibility on the theoretical level.

It should be noted that any chemical network in this paper is hypothetically shaped. With appropriate structure design, such an abstract set of reactions is said to be able to compute, or namely, simulate Markov processes. Nevertheless, some kind of physical substrate, such as DNAs or proteins, is required to emulate the system. In 2010, Soloveichik \cite{soloveichik2010dna} constructed systems of DNA molecules that could closely approximate the dynamic behavior of arbitrary uni- or bimolecular chemical networks, which endowed this purely conjectural computing method with meaningfulness. In this paper, an original DNA method is proposed, inspired by Soloveichik, for implementing any unimolecular network with only one product in each reaction. Bimolecular networks for second-order Markov chains are ought to be compiled to DNA strand displacement reactions as designed in article \cite{soloveichik2010dna}.

Notations in this paper are listed below for clearer reference.

\begin{table}[ht]\scriptsize
\caption{Notations in This Paper.}
\begin{tabular}{c|l||c|l}
\Xhline{1.2pt}
\cellcolor[gray]{0.8}\textbf{Symbol}&\multicolumn{1}{c||}{\cellcolor[gray]{0.8}\textbf{Definition}}&\cellcolor[gray]{0.8}\textbf{Symbol}&\multicolumn{1}{c}{\cellcolor[gray]{0.8}\textbf{Definition}}\\
\hline
\hline
$X_{S_j}(0)$ & number of molecules of molecular species $S_j$,& $k$ & reaction rate constant,\\
$X_{S_j}(t)$ & number of molecules of molecular species $S_j$ & $k_i$ & reaction rate constant of the $i$th reaction,\\
&at time $t$,& $c$ & reaction parameter,\\
$\bm{X}(0)$ & numbers of molecules of each species,& $c_i$ & reaction parameter of the $i$th reaction,\\
$\bm{X}(t)$ & numbers of molecules of each species at time $t$, & $\upsilon_i$ & the vector whose $j$th component is $\upsilon_{ji}$,\\
$x_{S_j}(0)$ & initial concentration of molecular species $S_j$,& $\upsilon_i'$ & the vector whose $j$th component is $\upsilon_{ji}'$,\\
$x_{S_j}(t)$ & concentration of molecular species $S_j$ at time $t$, & $\upsilon_{ji},\upsilon_{ji}'$ & nonnegative integers,\\
$\bm{x}(0)$ & initial concentration of each molecular species, & $Pr(A)$ & the probability of event $A$ occurring,\\
$\bm{x}(t)$ & concentration of each molecular species at time $t$ & $\mathcal{F}_t$ & the information about the system that is\\
$V$ & volume of the system,&& available at time $t$.\\
\Xhline{1.2pt}
\end{tabular}
\label{t1}
\end{table}

\section{Preliminaries}
Stochastic and deterministic models are two most common models for describing chemical reaction networks. Preliminaries are given below for preparing simulations and explaining the novelty of our work.
\subsection{Deterministic Model Versus Stochastic Model}
According to deterministic mass action kinetics \cite{anderson2011continuous,erdi1989mathematical,horn1972general}, a set of ODEs are derived to determine the concentration of each molecular type at transient time $t$ in the system. Species concentrations are solutions to ODEs, thus are continuous, single-valued functions of time. This model is also named as ordinary differential equation model. Generally, consider a network of $r_0$ reactions involving $s_0$ chemical species, $S_1,\dots,S_{s_0}$ in Eq. (\ref{eq:sum}), where $\upsilon_{ji},\upsilon_{ji}'$ are nonnegative integers.

\begin{equation}\label{eq:sum}
\sum_{j=1}^{s_0}\upsilon_{ji}S_j\to\sum_{j=1}^{s_0}\upsilon_{ji}'S_j, \quad i=1,\dots,r_0.
\end{equation}

ODEs in Eq. (\ref{eq:dbm}) are used to give the time evolution of the system. $k_i$ is the reaction rate constant of the $i$th reaction. $x_{S_j}(t)$ is the concentration of molecular species $S_j$ at time $t$. $x^{\upsilon_i}$ in Eq. (\ref{eq:dbm}) is defined in Eq. (\ref{eq:xup}). As soon as the ODEs are solved, the output of the chemical reaction system can be uniquely determined.

\begin{equation}\label{eq:dbm}
\frac{d\bm{x}(t)}{dt}=\sum_i k_ix^{\upsilon_i}(\upsilon_i'-\upsilon_i).
\end{equation}

\begin{equation}\label{eq:xup}
\begin{aligned}
x^{\upsilon_i}&\stackrel{def}{=}x_{S_1}(t)^{\upsilon_{1i}}\cdot x_{S_2}(t)^{\upsilon_{2i}}\cdots x_{S_{s_0}}(t)^{\upsilon_{s_0i}}\\
              &=\prod_{j=1}^{s_0}x_{S_j}(t)^{\upsilon_{ji}}.
\end{aligned}
\end{equation}


When it comes to stochastic models \cite{gillespie1976general,anderson2011continuous}, consider a very simple reaction: $A+B \xrightarrow{k} C$. Gillespie \cite{gillespie1976general} points out the probability that it will occur somewhere inside $V$ in the next infinitesimal time interval $\Delta t$ is given by: $cX_A(t)X_B(t)\Delta t$, where $c$ is reaction parameter and $k_i=Vc_i$. $X_A(t)$ stands for the molecule number of $A$ at time $t$ and $X_B(t)$ stands for that of $B$. Similarly, Anderson \cite{anderson2011continuous} assumes the same probability, taking no account of the volume of the system $V$, as in Eq. (\ref{eq:pre}). $\mathcal{F}_t$ is the condition of the available system information at time $t$. In Anderson's work, he also models the concentrations of a reaction network as complex random processes composed of Poisson processes.

\begin{equation}\label{eq:pre}
Pr\{\text{reaction occurs in }(t,t+\Delta t]|\mathcal{F}_t\} \approx kX_A(t)X_B(t)\Delta t.
\end{equation}

With this inherent random property of chemical reaction system, Gillespie puts forward a simulation algorithm based on Monte Carlo techniques. Note that for the Gillespie algorithm to be applicable, the number of reactant species for each reaction cannot exceed three.

\subsection{Comparison}
According to Gillespie\cite{gillespie1976general}, the mathematical relationship between $X_{s_i}$ and $x_{S_i}$ is that $x_{s_i}=X_{s_i}/V$, which is self-evident. Kurtz \cite{kurtz1972relationship} points out the relationship between the two models that in certain special cases and more complex systems, the deterministic model is the infinite volume limit of the stochastic one. This implies that the deterministic model is less accurate than the stochastic model when reactions occur in small compartments. The stochastic one takes account of fluctuations and correlations, providing better simulation for reality.



If expressed as a stochastic model as mentioned above, a chemical reaction network itself is a random process. When randomness is inherent to chemical reactions, the difference between building stochastic models for CRNs and molecular computation for stochastic problems needs illustrating in case of confusion. When building stochastic models, mathematical theories are utilized to analyze natural networks. In detail, the random variables are concentrations of each molecular type and there may be a multi-dimensional state space of different concentration values. When we solve stochastic problems using molecular reactions, these reactions are expected to express solutions in some way. For instance, in this paper, probability distributions are conveyed by concentrations. The random variables and state space depend on the particular case to be considered. The motivation as well as the Markov structure is entirely different. The comparison is summarized in Table \ref{t3}.

\begin{table}[ht]
\centering
\caption{Comparison between \emph{Building Stochastic Models for CRNs} and \emph{Molecular Computation for Stochastic Problems}.}
\begin{tabular}{c||m{5.5cm}<{\centering}|m{5.5cm}<{\centering}}
\Xhline{1.2pt}
\rowcolor[gray]{0.8}
\textbf{Aspects}&\textbf{Stochastic Models for CRNs}&\textbf{Molecular Computation for Stochastic Problems}\\
\hline
\hline
\textbf{Random Variables} & quantity of each molecule type & events\\
\hline
\textbf{State Space} & different molecule numbers & different possible events\\
\hline
\textbf{Simulation Model} & stochastic & stochastic or deterministic\\
\Xhline{1.2pt}
\end{tabular}
\label{t3}
\vspace*{4pt}
\end{table}

\section{Methodology}
\subsection{Discrete-Time Markov Chains}
Several essential concepts regarding Markov chains \cite{bolch2006queueing} need to be specified in the first place as follows.
\begin{Def}
A given stochastic process $\{X_0,X_1,\dots,$ $X_{n+1},$ $\dots\}$ at the consecutive points of observation $0,1,$ $\dots,$ $n+1,\dots$ constitutes a DTMC if the following relation, that is, the Markov property, holds for all $n \in \mathbb{N}_0$ and all $s_i \in S$:
\begin{equation}
\begin{aligned}
P(X_{n+1}&=s_{n+1}|X_n=s_n,X_{n-1}=s_{n-1},...,X_0=s_0)\\
             &=P(X_{n+1}=s_{n+1}|X_n=s_n).
\end{aligned}
\end{equation}
\end{Def}

In the homogeneous case, the transition probability from state $i$ to $j$ is independent of time $n$ and is defined as: $p_{ij}=P(X_{n+1}=j|X_n=i),~\forall n \in T$.The transition matrix $\textbf{P}=[p_{ij}]$.  Vector $\boldsymbol{\upsilon} (n)=({\upsilon}_0 (n),$ ${\upsilon}_1 (n),$ ${\upsilon}_2 (n),...)$ stands for the state probabilities at time $n$. The initial probability vector is $\boldsymbol{\upsilon} (0)=({\upsilon}_0 (0),$ ${\upsilon}_1 (0),$ ${\upsilon}_2 (0),...)$. As $n \to \infty$, the probability vector $\boldsymbol{\upsilon} (n)$ that converges is called steady state probability vector.


\subsubsection{Example}
Consider such a gambler's ruin problem \cite{degrootprobability} referred as the probability of winning in an unfavorable game. Suppose that the probability that gambler $A$ will win one dollar on any given play is $0.4$. Suppose also that the initial fortune of gambler $A$ is $9$ dollars and the initial fortune of gambler $B$ is just one dollar. We need to determine the probability $a_i$ that gambler $A$ wins one dollar from gambler $B$ before gambler $B$ wins $9$ dollars from gambler $A$.

The required probability $a_i$ is given by Eq. (\ref{eq:ai}) through mathematical analysis:
\begin{equation}\label{eq:ai}
a_i=\frac{(\frac{3}{2})^{9}-1}{(\frac{3}{2})^{10}-1} \approx \frac{2}{3}=0.66.
\end{equation}

\begin{figure}[htbp]
\centering
\includegraphics[width=8cm]{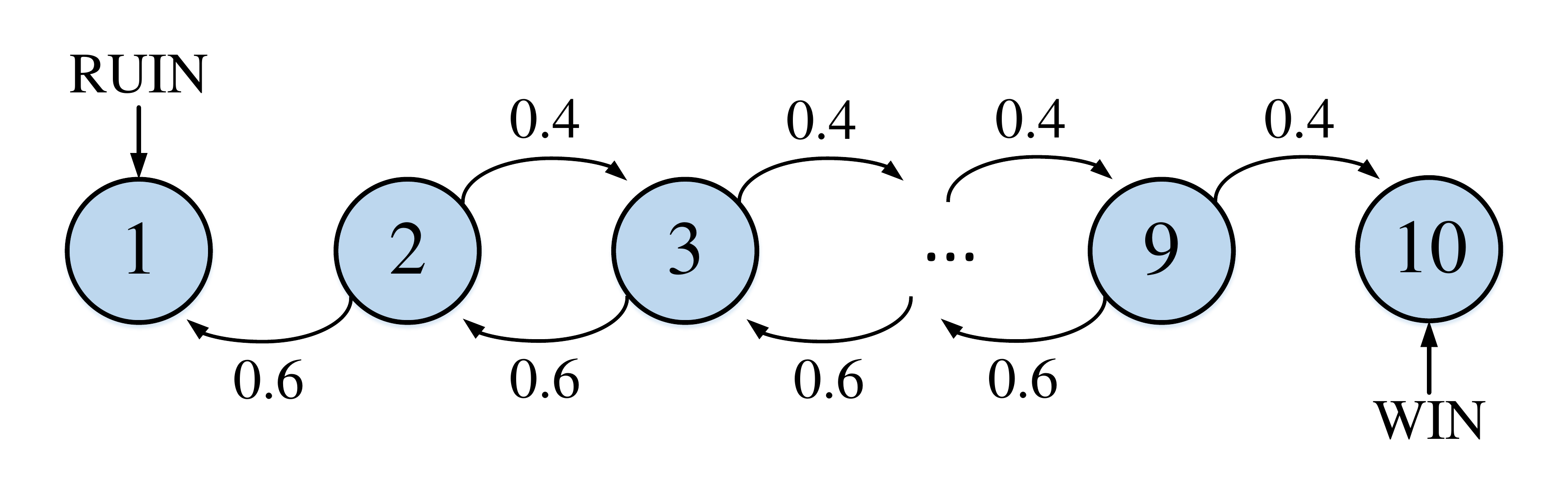}
\caption{The probability of winning in an unfavorable game.}\label{fig:gambler}
\end{figure}

This problem is considered as a DTMC illustrated in Fig. \ref{fig:gambler}. There are $10$ states $\upsilon_1,\upsilon_2,\dots,\upsilon_{10}$, with $\upsilon_i$ indicating that gambler $A$ holds $i$ dollars. From the state transition diagram, states $\upsilon_2,\upsilon_3,\dots,$ $\upsilon_{9}$ may jump to the previous or next state, while states $\upsilon_1$ and $\upsilon_{10}$ are absorbing. Transition probabilities are given by the probabilities of winning and losing. Utilizing this transition property, we try to model it by a chemical reaction network, with each molecular species $\upsilon_i$ representing each state $\upsilon_i$ as firstly proposed by Salehi \cite{salehi2015markov}. In that each reaction has a rate constant $k$ and this $k$ influences the probability of reaction occurring to some degree as in Eq. (\ref{eq:pre}), we endeavour to harness this attribute and map Markov chain's transition probabilities to reaction rate constants. Then the one-to-one correspondence between state transitions and reactions is established. The finished network is shown in Eq. (\ref{eq:99}).

\begin{equation}\label{eq:99}
\left\{
 \begin{aligned}
&\upsilon_2 \xrightarrow{0.6} \upsilon_1, \ \upsilon_3 \overset{0.6}{\underset{0.4}{\rightleftharpoons}} \upsilon_2,\\
&\dots\\
&\upsilon_{9} \overset{0.6}{\underset{0.4}{\rightleftharpoons}} \upsilon_{8},\ \upsilon_{9} \xrightarrow{0.4} \upsilon_{10}.
\end{aligned}
\right.
\end{equation}

To obtain the steady state distribution, the Markov chain needs to be assigned an initial distribution. As molecular concentrations are desired to be the system's indicators, the concentration of each molecular species is initialized with the corresponding state's initial probability. In this problem, gambler $A$ holds $9$ dollars in the beginning, therefore $\boldsymbol{\upsilon} (0)=(0,0,0,0,0,0,0,0,1,0)$. Accordingly, $\bm{x}(0)=(0,0,0,0,0,0,0,0,1,0)$. When well prepared, the system begins to react and all required to be done is waiting until it reaches an equilibrium state, helping processing the computation with the chemical potential energy. Then the final concentrations are the outputs: steady state probabilities. Simulations will be given.

\begin{Rem}
As is known, any chemical reaction network itself is a Markov chain, thus it is easily misunderstood that the mapping is a self-existed conclusion. Nevertheless, constructing Markov chain model for a chemical system, the state space is usually determined by concentrations or molecule numbers instead of molecule species. For example, consider the reaction $A \rightleftharpoons B$ and suppose that initially there are two molecules in total and they can be either $A$ or $B$. The transition is mutual but the overall molecule number remains two. Obviously, $X_A(t)$ or $X_B(t)$ is a Markov chain with state space $\{0,1,2\}$. $\{X_A(t),X_B(t)\}$ can also be a Markov chain with state space $\{\{0,2\},\{1,1\},\{2,0\}\}$. In real DNA reactions, the concentration is always scaled to $nM$ or $M$, which means the molecular number is much larger than two. Therefore, the Markov chain model for Eq. (\ref{eq:99}) is apparently not that in Fig. \ref{fig:gambler}. The delicately designed structure of this network, happens to be capable of modeling this chain's computation when on a large molecular scale and giving a relatively deterministic result.
\end{Rem}

\subsubsection{Design Concept}
\begin{figure}[ht]
\centering
\includegraphics[width=14cm]{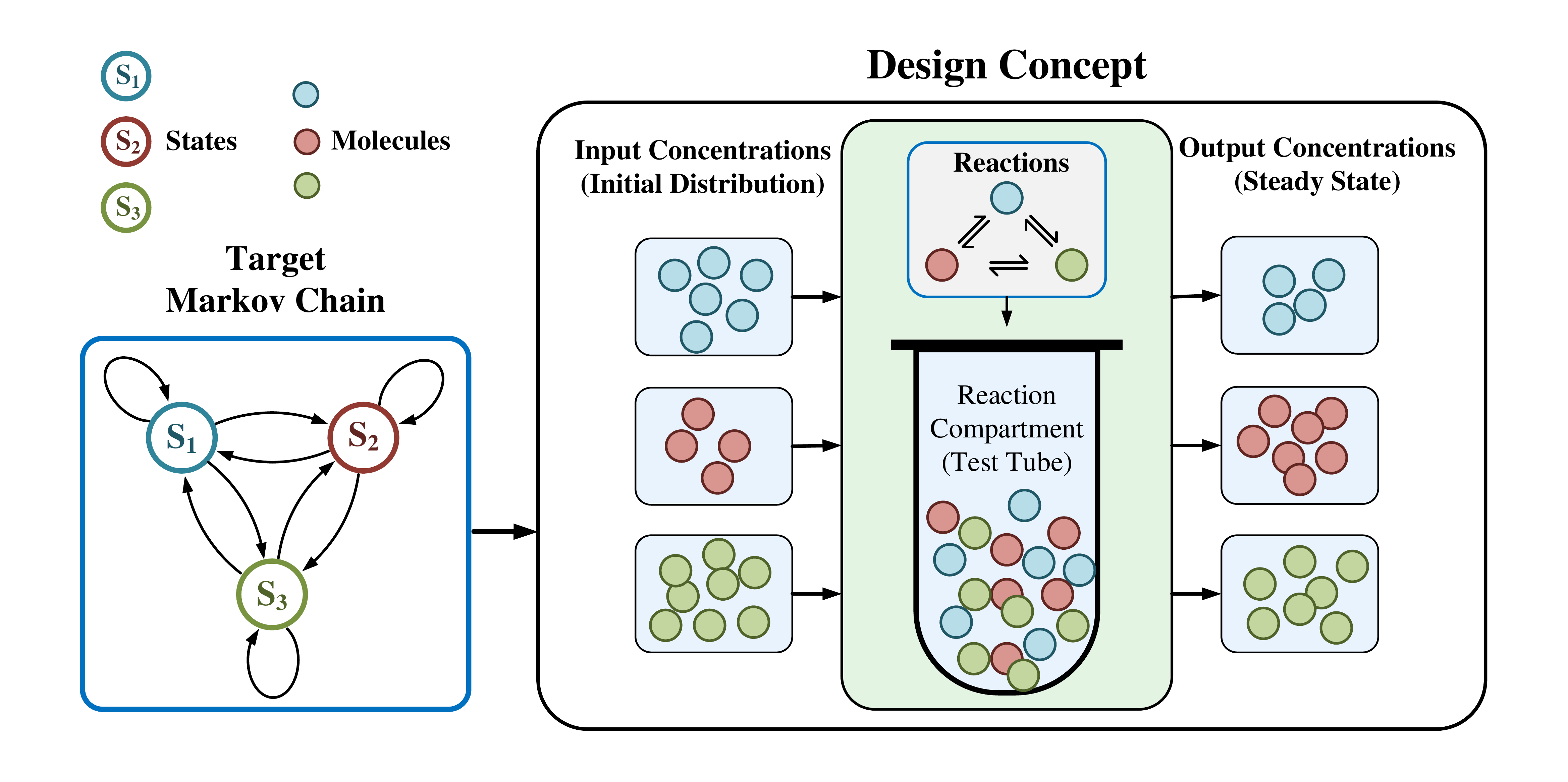}
\caption{Framework for the design concept.}\label{fig:model}
\end{figure}

As the gambler's ruin problem is explained in detail, the entire design approach gradually becomes clear and easy to understand. The conclusive framework for the design concept is depicted in Fig. \ref{fig:model}. Given a target Markov chain, each state is modeled by a unique molecular type. Unimolecular reactions are constructed to implement state transitions, with one type of molecule changing into another. Input concentrations are initialized to activate the system and output concentrations provide expected stationary distribution.

The fact that jumps between two states may exist at the same time causes the derived reactions to be reversible. Given that there are $k$ states in all, the reactions to implement this target DTMC are shown in Eq. (\ref{eq:vi}).
\begin{equation}\label{eq:vi}
\upsilon_i \overset{k_{ij}}{\underset{k_{ji}}{\rightleftharpoons}} \upsilon_j,~ i=1,2,...,k,~j=1,2,...,k,~i \ne j.
\end{equation}

In summary, the complete method includes $6$ steps: \emph{\textbf{Step 1)}} Model each state $\upsilon_i$ by a molecular type $\upsilon_i$. \emph{\textbf{Step 2)}} Model each transition probability $p_{ij}$ by rate constant $k_{ij}$. \emph{\textbf{Step 3)}} Model all state transitions by reactions $\upsilon_i \overset{k_{ij}}{\underset{k_{ji}}{\rightleftharpoons}} \upsilon_j$. \emph{\textbf{Step 4)}} Set the values of rate constants $k_{ij}$ proportional to the corresponding transition probabilities $p_{ij}$. \emph{\textbf{Step 5)}} Set the initial concentrations of molecular types $\upsilon_i$ according to probability $\upsilon_i(0)$. \emph{\textbf{Step 6)}} If the steady state solution exists, compute the steady state solution of the DTMC by the final concentration of species $\upsilon_i$.

\textbf{Complexity Analysis:} In our approach, control molecules are not required compared to Salehi's \cite{salehi2015molecular}, thus the cost of molecular types is saved. Since each state is represented by one unique molecular type, the number of molecular types equals the number of states $k$. The number of reactions depends on the particular case. Specifically, if the transitions between two states exist at the same time, one reversible reaction functions to realize them. As one reaction can compute either one transition or two transitions, the number of reactions equals or is smaller than the number of all transitions $m$. If there are no mutual transitions, all reactions are not reversible and the reaction number meets its maximum value $m$.

\subsection{Continuous-Time Markov Chains}
After the DTMC is well synthesized, it is excitingly found that the approach can also be extended to implement CTMC. Some detailed mathematical descriptions are provided here for further formal analysis.

\begin{Def}
A given stochastic process $\lbrace X_t : t\in T \rbrace$ constitutes a \emph{continuous-time Markov chain} if for arbitrary $t_i \in \mathbb{R}_0^+$, with $0=t_0<t_1 < \cdots <t_n<t_{n+1}$, $\forall n \in \mathbb{N}$, and $\forall s_i \in S$ (state space of this chain), the following relation holds:
\begin{equation}
\begin{aligned}
P(X_{t_{n+1}}&=s_{n+1}|X_{t_n}=s_n,X_{t_{n-1}}=s_{n-1},...,X_{t_0}=s_0)\\
             &=P(X_{t_{n+1}}=s_{n+1}|X_{t_n}=s_n).
\end{aligned}
\end{equation}
\end{Def}

${\pi}_i (u)$ stands for the probability of state $i$ at any instant of time $u$. Vector $\boldsymbol{\pi} (u)=({\pi}_0 (u),{\pi}_1 (u),$ ${\pi}_2 (u),...)$ stands for the state probabilities at any instant of time $u$. Unlike the discrete-time case, the state probabilities of a CTMC cannot be computed easily by transition probabilities. Therefore, we define the instantaneous transition rates $q_{ij}$ ($i\ne j$) of the CTMC traveling from state $i$ to state $j$. For all states with $i\ne j$, we define $q_{ij}(t)=\lim_{\Delta t \to 0} \frac{p_{ij}(t,t+\Delta t)}{\Delta t}~(i \ne j), q_{ii}(t)=\lim_{\Delta t \to 0} \frac{p_{ii}(t,t+\Delta t-1)}{\Delta t}.$ If the limits do exist, it is clear that at any instant of time $t$, the following reaction holds:
\begin{equation}\label{eq:qii}
\sum_{j \in S} q_{ij}(t)=0, \forall i \in S.
\end{equation}

Eq. (\ref{eq:qii}) can be modified as:
\begin{equation}
q_{ii}(t)=-\sum\nolimits_{j,j \ne i}q_{ij}(t).
\end{equation}

In the time-homogeneous case, with time-independent transition rates $q_{ij}=q_{ij}(t), \forall i,j \in S$ and the system of differential Eq. (\ref{eq:dif}), we describe a CTMC:
\begin{equation}\label{eq:dif}
\frac{d {\pi}_j(t)}{dt}=\sum_{i \in S}q_{ij} {\pi}_i (t), \forall j \in S.
\end{equation}

The infinitesimal generator matrix $\textbf{Q}=[q_{ij}],~\forall i,j \in S$. If existing for a given CTMC, the steady state probabilities are independent of time and we immediately get: $\lim_{t \to \infty}\frac{d \boldsymbol{\pi}(t)}{dt}=0.$

As specified by the mathematical definition, the only difference between CTMC and DTMC is that for any instant of time, CTMC has a probability distribution $\boldsymbol{\pi} (u)$, which is called the transient solution. The state space is still discrete here. The proposed method will be able to compute not only the steady state distribution but also the transient solution of an arbitrary CTMC.

\subsubsection{Example}
Two common cases in queueing theory \cite{bolch2006queueing} are used to exemplify the computation of transient solution and steady state solution, respectively.

\begin{figure}[htbp]
\centering
\includegraphics[width=8cm]{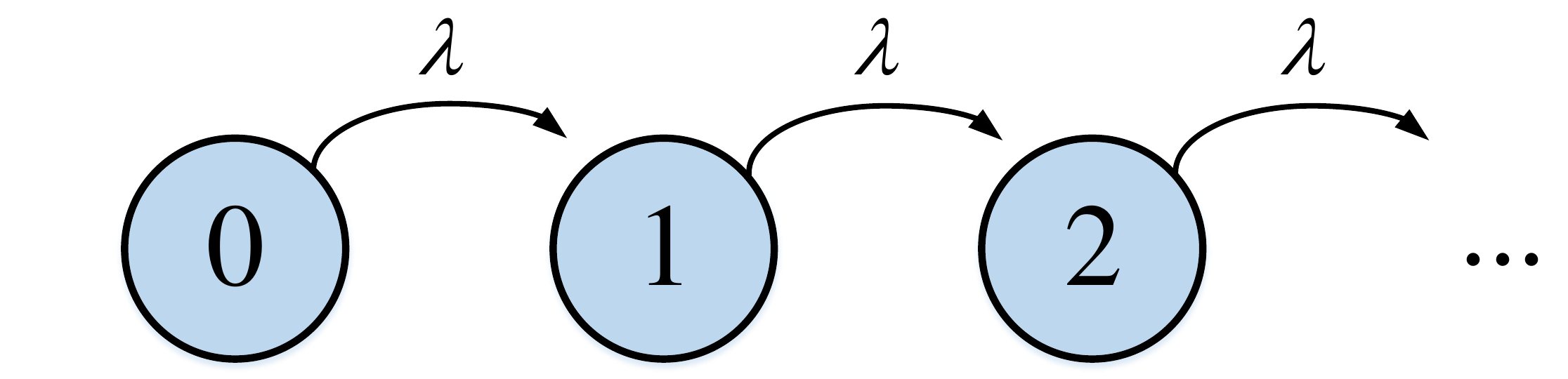}
\caption{A pure birth process.}\label{fig:purebirth}
\end{figure}

\textbf{A Pure Birth Process:} Consider the infinite state CTMC depicted in Fig. \ref{fig:purebirth} representing a pure birth process with constant birth rate $\lambda$. The only possible transitions are from state $k$ to state $k+1$ with rate $\lambda$. Note that this is a nonirreducible Markov chain for any finite value of $\lambda$, so the steady state solution does not exist.

According to the transition graph, the only difference from DTMC is that the transition rate is no longer a value of probability, while the structure is analogous. Hence, the technique can continue to be used, with rate constants modeling transition rates as opposed to transition probabilities. Given that only finite states are feasible in CRNs, we implement this CTMC with a $6$-state one, the first $5$ transient solutions of which are exactly the same as the pure birth process. The reactions are presented in Eq. (\ref{eq:purerea}). The setting of initial concentrations follows the same principle as DTMC. Here, the time evolution of concentrations ideally resembles the transient solutions. Please refer to simulation for details.
\begin{equation}\label{eq:purerea}
\pi_0\xrightarrow{\lambda} \pi_1,~
\pi_1\xrightarrow{\lambda} \pi_2,~
\dots~
\pi_4\xrightarrow{\lambda} \pi_5.
\end{equation}

\begin{figure}[htbp]
\centering
\includegraphics[width=7.8cm]{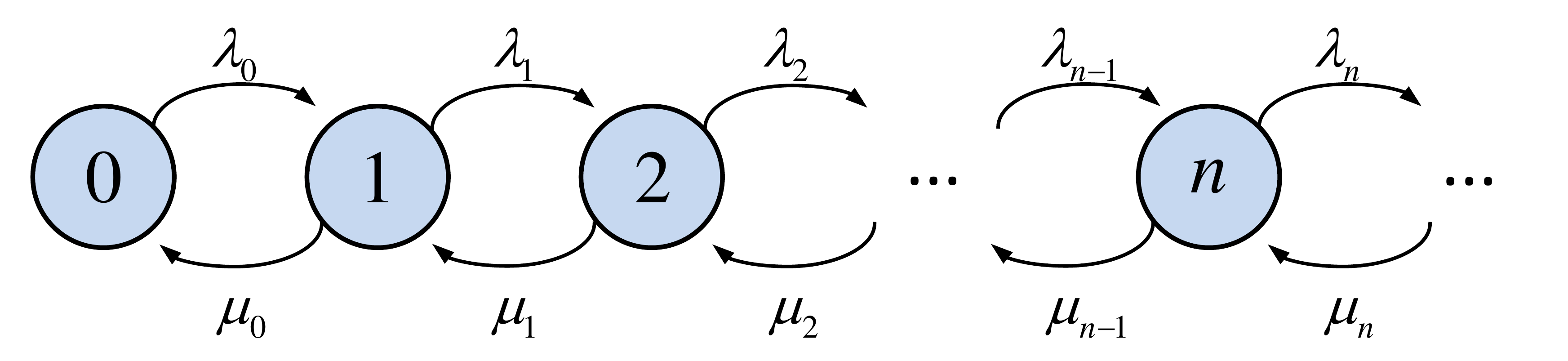}
\caption{A birth-death process.}\label{fig:birth}
\end{figure}

\textbf{A Birth and Death Process:} When it comes to steady state probabilities, another example serves better to verify our approach. A birth and death process is a Markov chain where transitions are allowed only between neighboring states. A one-dimensional birth-death process is shown in Fig. \ref{fig:birth}. In particular, a birth-death process with a constant birth rate $\lambda$ (arrival rate) and a constant death rate $\mu$ (service rate) is called an $M\backslash M\backslash 1$ Queue. This case is used to illustrate how to compute both transient and stationary solutions.


Similarly, only finite states are realizable in CRNs thus we could only get the approximate solutions. To implement such a CTMC, we have the reactions designed in Eq. (\ref{eq:death}).

\begin{equation}\label{eq:death}
\pi_0\overset{\lambda}{\underset{\mu}{\rightleftharpoons}} \pi_1,~
\pi_1\overset{\lambda}{\underset{\mu}{\rightleftharpoons}} \pi_2,~
\dots~
\pi_4\overset{\lambda}{\underset{\mu}{\rightleftharpoons}} \pi_5.
\end{equation}

\subsubsection{Design Concept}
As clarified by the two cases, it is only slightly different to implement a CTMC. The transition rates instead of probabilities are modeled by the rate constants. The designed reactions for implementing an arbitrary CTMC are shown in Eq. (\ref{eq:pi})
\begin{equation}\label{eq:pi}
\pi_i \overset{k_{ij}}{\underset{k_{ji}}{\rightleftharpoons}} \pi_j,~i=1,2,...,k,~j=1,2,...,k,~i \ne j.
\end{equation}

Such mapping can, to a great extent, model and implement the CTMC, computing not only the steady state probabilities but also the transient solutions. More specifically, the concentrations of the molecules $\pi_i$ at any instant time of $t$ are the same as the probability distribution of the CTMC at time $t$. In addition, the final concentrations of $\pi_i$ are the steady state probabilities of the target CTMC. To sum up, the entire procedure contains $7$ steps: \emph{\textbf{Step 1)}} Model each state $\pi_i$ by a molecular type $\pi_i$. \emph{\textbf{Step 2)}} Model each transition rate $q_{ij}$ by reaction rate constant $k_{ij}$. \emph{\textbf{Step 3)}} Model all state transitions by reactions $\pi_i \overset{k_{ij}}{\underset{k_{ji}}{\rightleftharpoons}} \pi_j$ when $q_{ij}>0$. \emph{\textbf{Step 4)}} Set the values of rate constants $k_{ij}$ proportional to the transition rates $q_{ij}$. \emph{\textbf{Step 5)}} Set the initial concentrations of molecular types $\pi_i$ according to probability $\pi_i(0)$. \emph{\textbf{Step 6)}} Compute the transient solution of the CTMC by the concentrations of $\pi_i$ at any instant of time $t$. \emph{\textbf{Step 7)}} If the steady state solution exists, compute the steady state solution of the CTMC by the equilibrium concentrations of $\pi_i$. The complexity is the same as DTMC thus omitted here.

\subsubsection{ODE Analysis}
The correctness of our methodology could be predicted to a great extent by simulation results. However, beyond simulation, the congruence between the ODE model of our designed network and that of the corresponding CTMC can prove the validity in a mathematical way.

\begin{proof}
According to deterministic mass action, the ODEs of Eq. (\ref{eq:pi})'s network can be simply derived in Eq. (\ref{eq:fra}).
\begin{equation}\label{eq:fra}
\frac{dx_{\pi_i}(t)}{dt}=\sum\nolimits_{j,j\ne i}k_{ji}x_{\pi_j}(t)-x_{\pi_i}(t)\sum\nolimits_{j,j\ne i}k_{ij},~i=1,2,...,k.
\end{equation}

According to Eq. (\ref{eq:dif}), the differential system to describe the target Markov chain is:
\begin{equation}\label{eq:pii}
\begin{aligned}
\frac{d\pi_i(t)}{dt}&=\sum_{j=1}^k \pi_i(t)q_{ji}\\
                    &=\sum\nolimits_{j,j\ne i} \pi_i(t)q_{ji}+\pi_i(t)q_{ii},~i=1,2,...,k.
\end{aligned}
\end{equation}

Bringing Eq. (\ref{eq:qii}) into Eq. (\ref{eq:pii}), we have:
\begin{equation}\label{eq:dpi}
\frac{d\pi_i(t)}{dt}=\sum\nolimits_{j,j\ne i} q_{ji}\pi_i(t)-\pi_i(t)\sum\nolimits_{j,j\ne i}q_{ij},~i=1,2,...,k.
\end{equation}

It can be found that the form of Eq. (\ref{eq:fra}) and Eq. (\ref{eq:dpi}) mirrors each other, thus proving that the solutions of the designed CRN are the same as the transient solutions of the CTMC. Therefore, the time evolution of concentrations can well reflect the transient probabilities at any instant of time. In addition, we define the final concentration of a given molecular type as the concentration of it when $t$ verges to $\infty $. And as such, the final concentrations of $\pi_i$ are the steady state probabilities of the target CTMC.
\end{proof}


\subsection{Two-Order Markov Chains}
In the previous sections, the Markov processes' transition probabilities depend only on the current state. Such chains are called first-order Markov chains. For the higher-order Markov chains, the transition probabilities depend on the current state and some previous states\cite{ching2013higher}. In point of fact, any n-order Markov chain can be expressed as a first-order chain with state space $S^n$, where $S$ is the state space of the original chain. Consequently, higher-order Markov chains can be implemented by the approach specified above. Unfortunately, this would exponentially increase the complexity. This problem may be resolved by increasing the dimension of CRNs instead of state space. However, there would exist a trade off between complexity and accuracy. Here we make use of bimolecular reactions to implement second-order Markov chains, where the transition probabilities depend on the latest two states---the current state and the previous state as shown in Eq. (\ref{eq:higher}).
\begin{equation}\label{eq:higher}
\begin{aligned}
P(X_{n+1}&=s_{n+1}|X_n=s_n,X_{n-1}=s_{n-1},...,X_0=s_0)\\
             &=P(X_{n+1}=s_{n+1}|X_n=s_n,X_{n-1}=s_{n-1}).
\end{aligned}
\end{equation}

\subsubsection{Example}
Higher-order Markov chains are usually used to predict weather because the future weather trend considerably depends on the previous records. Make allowance for such a simple model: tomorrow's weather depends on today and yesterday. Transition probabilities are given in Eq. (\ref{eq:0.9}), where $d_1,d_2,d_3$ represent day$1$, day$2$, day$3$ and $S,R$ represent sunny and rainy.

\begin{figure}
\centering
\includegraphics[width=12cm]{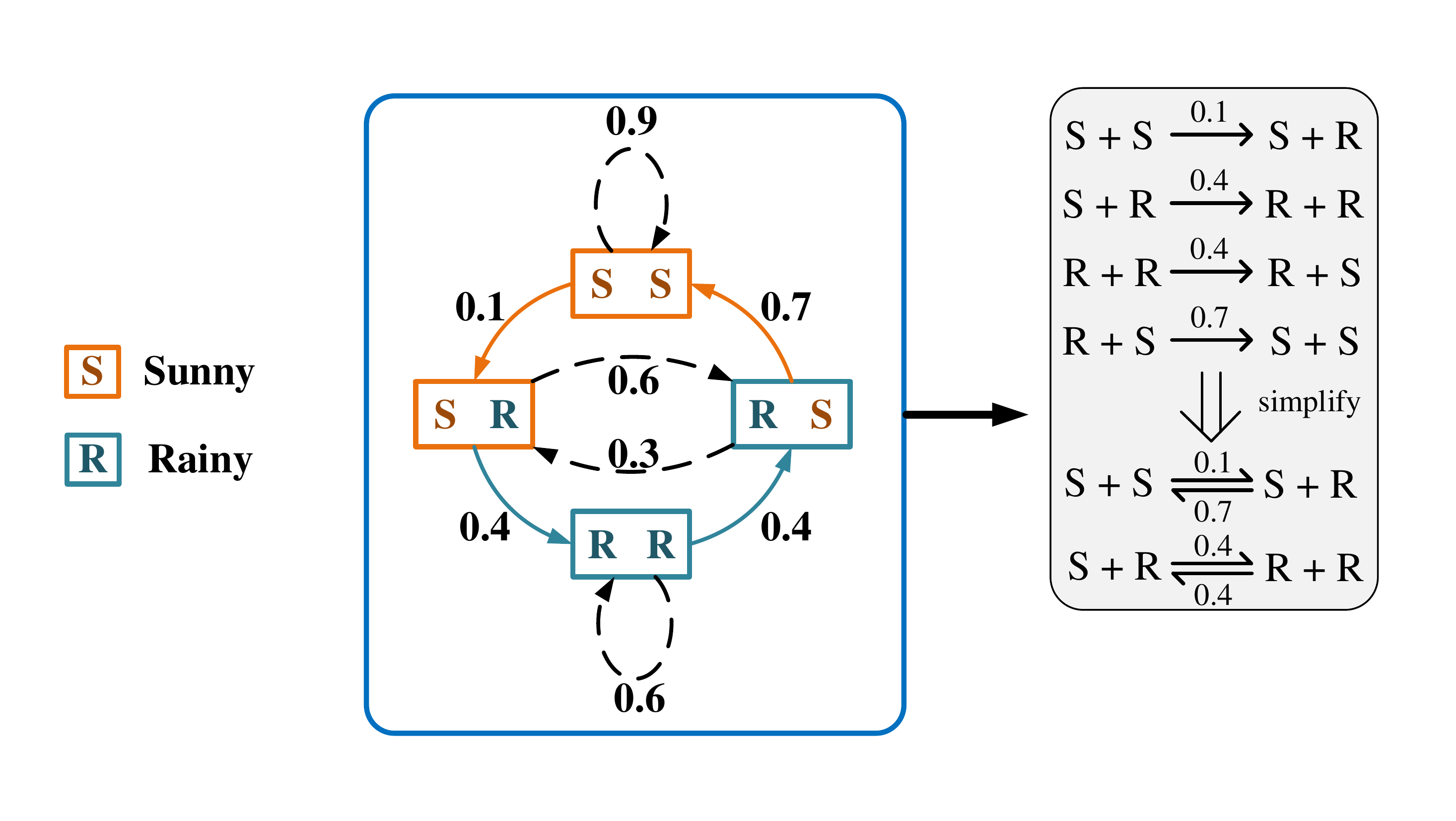}
\caption{Model for weather prediction.}\label{fig:weather}
\end{figure}

The state space is $\{S,R\}$, clearly. As shown, if the first day and the second day are both sunny, there is a $90\%$ chance that the third day is continuously sunny. If expressed as a first-order Markov chain, the state space will become $\{\{S,S\},\{S,R\},\{R,S\},\{R,R\}\}$ and the state transition diagram can be derived as in Fig. \ref{fig:weather}.

\begin{equation}\label{eq:0.9}
\resizebox{.5\hsize}{!}{$
p(d_3|d_2,d_1)=\left\{
\begin{aligned}
0.9,\text{ if } d_3=S,d_2=S \text{ and } d_1=S\\
0.1,\text{ if } d_3=R,d_2=S \text{ and } d_1=S\\
0.7,\text{ if } d_3=S,d_2=S \text{ and } d_1=R\\
0.3,\text{ if } d_3=R,d_2=S \text{ and } d_1=R\\
0.6,\text{ if } d_3=S,d_2=R \text{ and } d_1=S\\
0.4,\text{ if } d_3=R,d_2=R \text{ and } d_1=S\\
0.4,\text{ if } d_3=S,d_2=R \text{ and } d_1=R\\
0.6,\text{ if } d_3=R,d_2=R \text{ and } d_1=R\\
\end{aligned}
\right.
$}
\end{equation}

Instead of modeling each node by one molecular type, we model each state by one molecular type just as we do previously, so only two types of molecule are needed in all. To map each transition into one chemical reaction, it is found from the diagram that either node before or after the transition contains two states, resulting in the reactions' being bimolecular as shown in the right part of Fig. \ref{fig:weather}. For each reaction, reactants reflect the two previous states before transition and the two products are states after transition. Four transitions are drawn in dashed lines because the composition of states does not change, unable to form a new reaction. Finally, four irreversible reactions are derived and then simplified into two reversible reactions. The value assignment of rate constants and initial concentrations follows the same principle as in first-order chains. The equilibrium concentrations are expected to compute the stationary distribution and help predict the weather in the long run.

\subsubsection{Design Concept}
\begin{figure}[htbp]
\centering
\includegraphics[width=14cm]{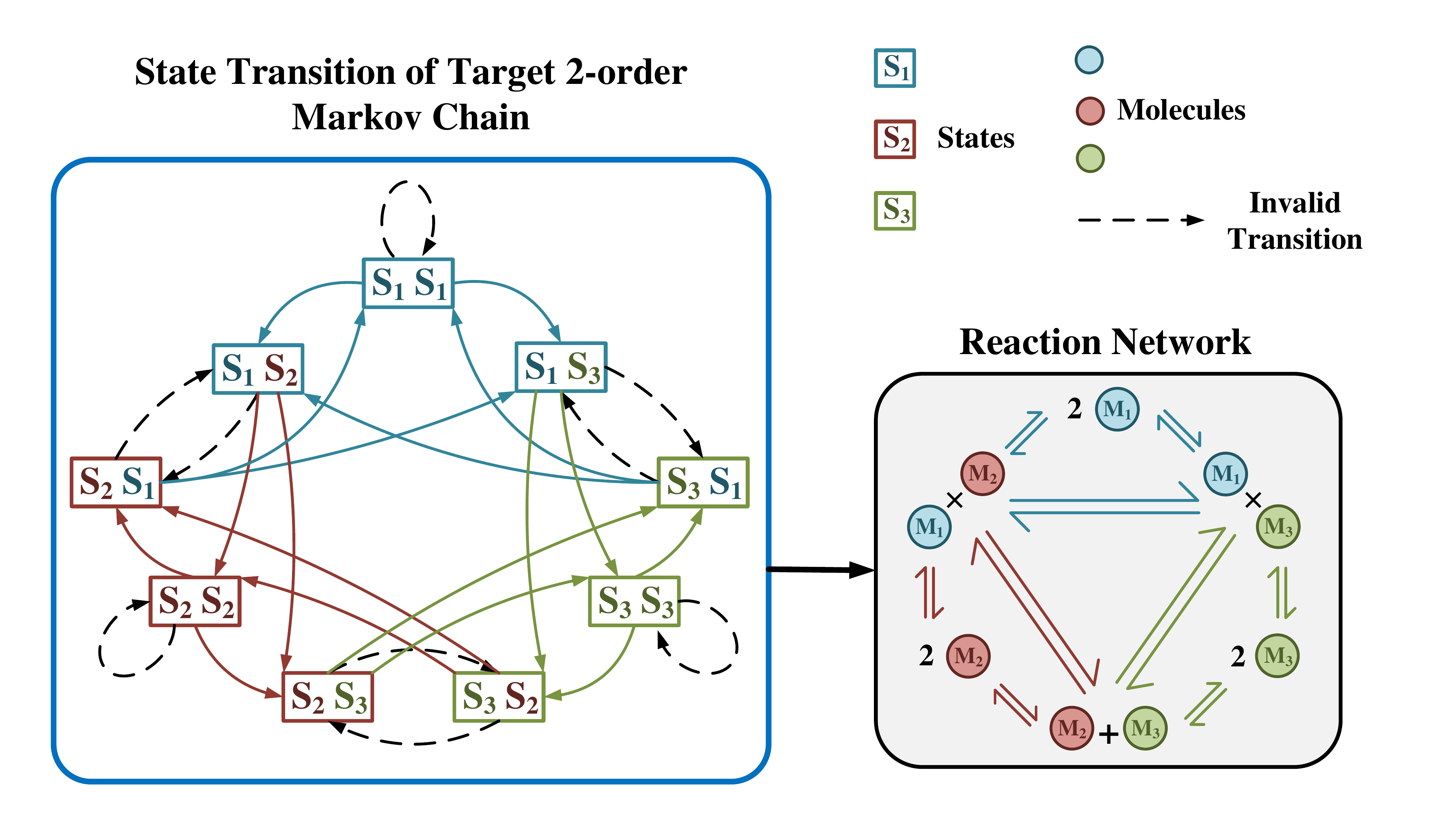}
\caption{Framework for design concept of second-order Markov chains.}\label{fig:secondmodel}
\end{figure}

Utilizing a $3$-state diagram, the framework for our design concept is summarized in Fig. \ref{fig:secondmodel}. Each transition produces one corresponding reaction, which has two reactants and two products. The reactants and products share one common molecule. Such implementation encourages the transition into a new state based on the two previous states, with the justified probability. Invalid transitions are unable to add reactions to the network.

The final approach for second-order Markov chains is concluded by $7$ steps: \emph{\textbf{Step 1)}} Model each state by a molecular type. \emph{\textbf{Step 2)}} Model each transition probability by one rate constant. \emph{\textbf{Step 3)}} Model each state transition by one bimolecular reaction. \emph{\textbf{Step 4)}} Exclude invalid transitions. \emph{\textbf{Step 5)}} Set the values of rate constants proportional to the transition probabilities. \emph{\textbf{Step 6)}} Set the initial concentrations of molecular types according to initial probabilities. \emph{\textbf{Step 7)}} If the steady state solution exists, compute the steady state solution of the second-order Markov chain by the equilibrium concentrations of the corresponding molecules.

\textbf{Complexity Analysis:} If there are $k$ states in all, $k$ molecular types are needed. The number of reactions equals or is smaller than the number of valid transitions $m$. When all transitions are possible, $m$ reaches its maximal value $k^3-k^2$. If the second-order chain is implemented by the approach of first-order chain, $k^2$ molecular types are required and $m$'s maximal value becomes $k^3-k$. Hence, this approach is increasingly efficient as $k$ rises.

\section{Simulation Results}
In order to ensure the DNA implementation is applicable \cite{soloveichik2010dna}, two constraints should be satisfied: maximal second-order rate constants are about $10^6$/M/s; maximum concentrations are on the order of $10^{-5}$M.
\subsection{Deterministic Simulation}
Since dynamics of a CRN endowed with mass action kinetics can be well demonstrated by ODEs, ODE-based simulation is usually a good solution to synthesize a CRN, offering a smooth output graph.

\begin{figure}[htbp]
\centering
\includegraphics[width=8cm]{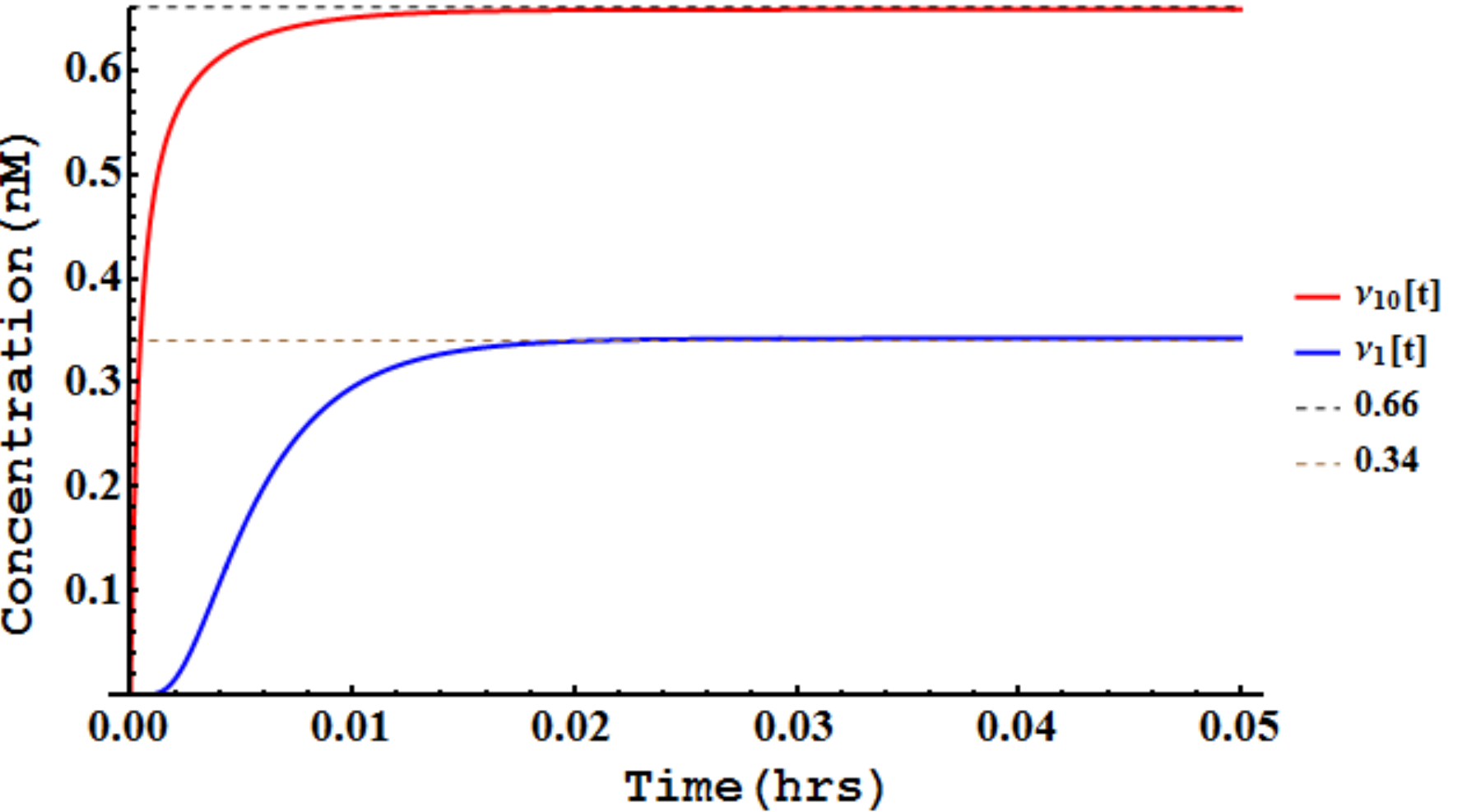}
\caption{ODE simulation result for gambler's ruin problem.}\label{fig:gamblerode}
\end{figure}

\textbf{Gambler's Ruin Problem:} As designed above, the unscaled rate constants are $0.4$ and $0.6$ and the unscaled initial concentration is $1$. To add feasibility, the scaled rates are chosen to be $0.4$/s and $0.6$/s here and the scaled concentration is $10^{-9}$M. The simulation result is shown in Fig. \ref{fig:gamblerode}.

From the graph, the concentrations approximate the accurate result $0.66$ and $0.34$ along the time line, meaning the probability of gambler $A$ winning one dollar ends up $0.66$. After $0.05$ hours, the error is less than $0.34\%$.

\begin{figure}[htbp]
\centering
\includegraphics[width=8cm]{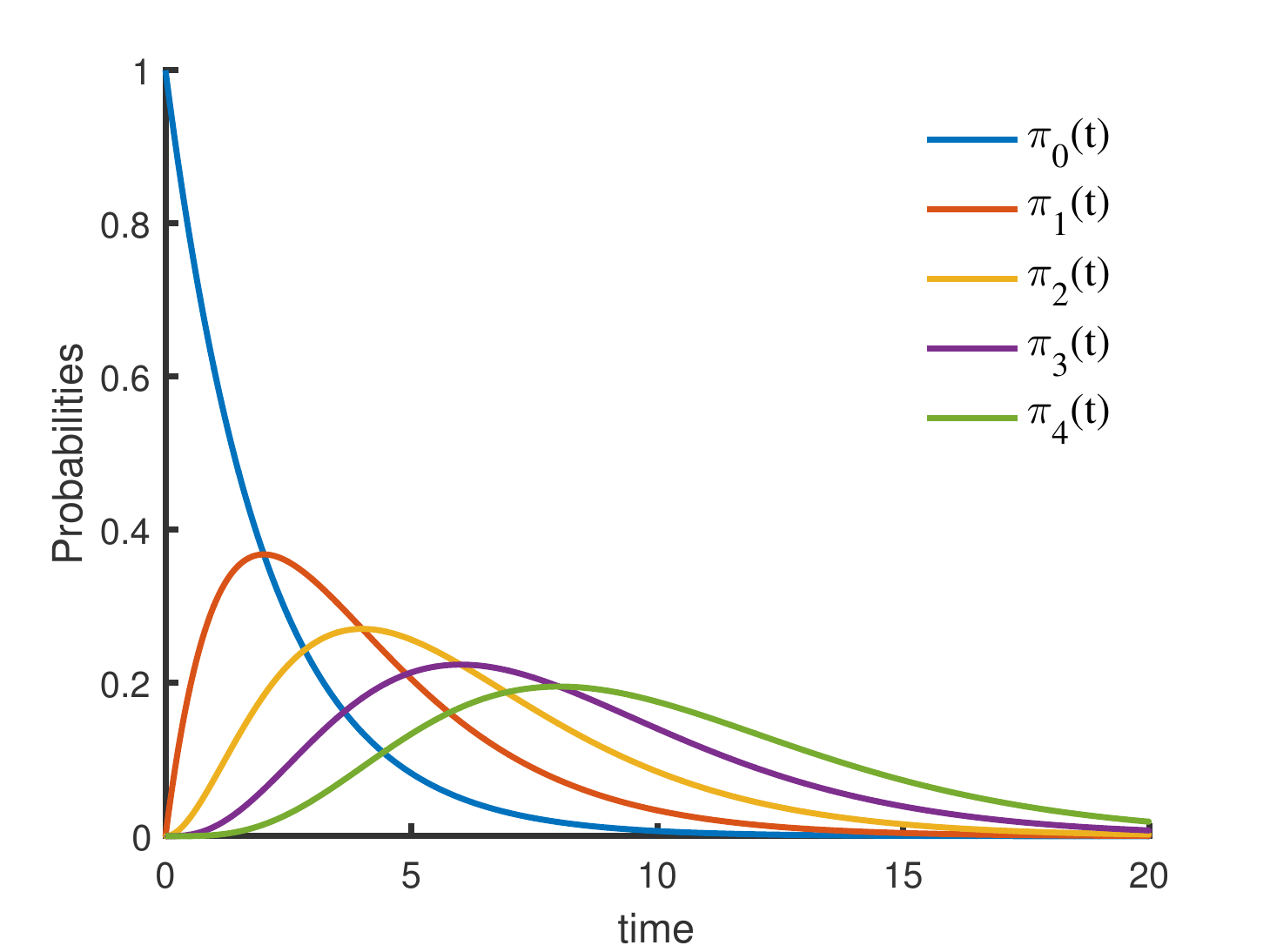}
\caption{Selected transient solutions for a pure birth process with $\lambda=0.5$.}\label{fig:figure2}
\end{figure}

\textbf{Pure Birth Process:} If initial probabilities $\pi_0(0)=1$ and $\pi_k(0)=0$ for $k\ge 1$, we can get a closed-form solution for each transient state probability $\pi_k(t)=\frac{{(\lambda t)}^k}{k!}e^{-\lambda t},k\ge 0$. If the parameter $\lambda$ is considered to be $0.5$, the graph is easily obtained by means of MATLAB as depicted in Fig. \ref{fig:figure2}.

\begin{figure}[htbp]
\centering
\includegraphics[width=8cm]{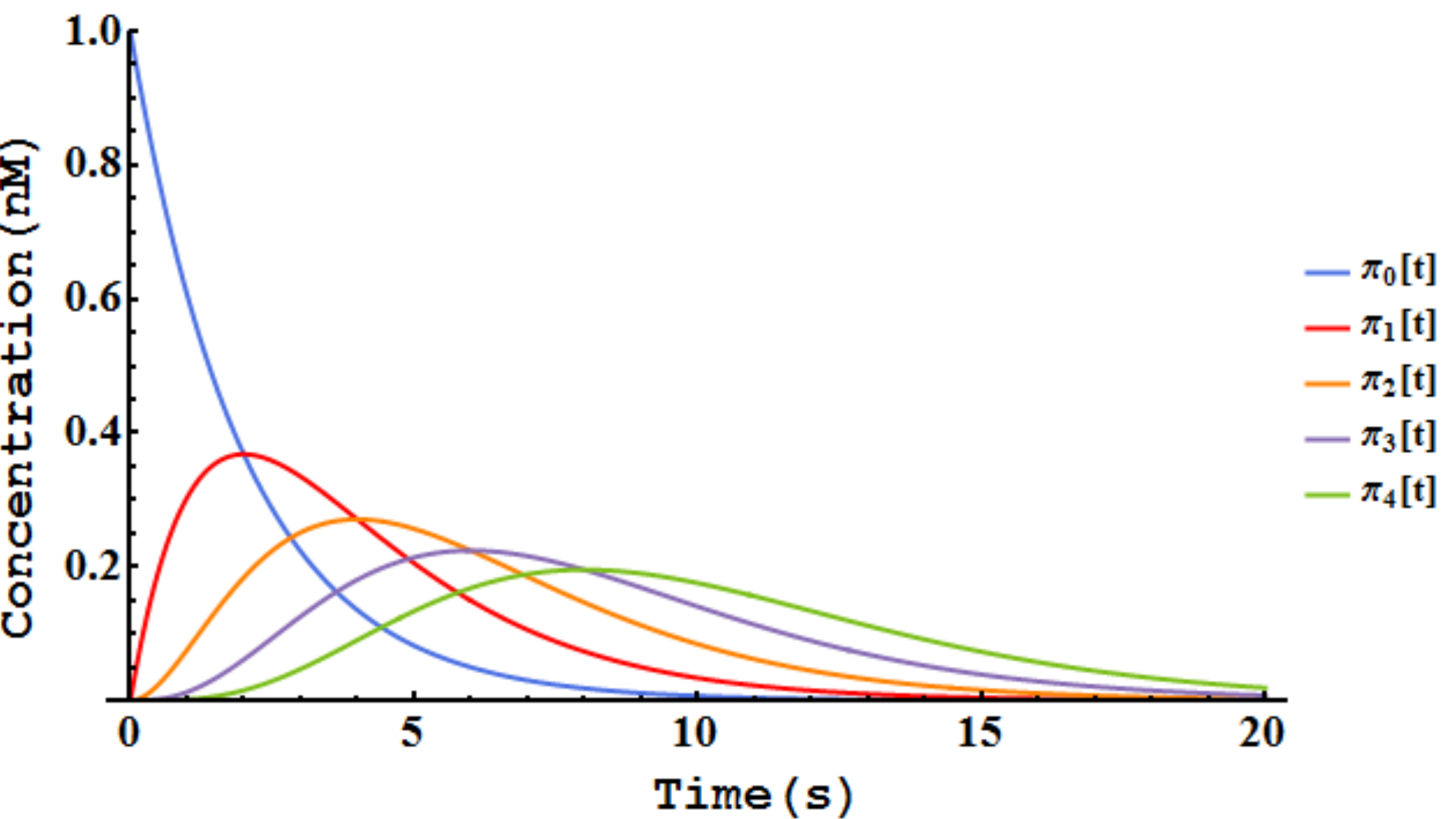}
\caption{ODE simulation result for a pure birth process with $\lambda=0.5$.}\label{fig:figure4}
\end{figure}

The initial concentration of $\pi_0$ is unscaled $1$ and those of other molecules are $0$. All the rate constants are unscaled $0.5$ when $\lambda=0.5$. The scaled concentration and rate are $10^{-9}$M and $0.5$/s. Simulation result is illustrated in Fig. \ref{fig:figure4}. Comparing the transient solution graphs of CTMC and the simulation graphs of CRNs, they resemble each other perfectly, realizing the desired functionality smoothly.

\textbf{Birth and Death Process:} With the initial state probabilities $\pi_0(0)=1$ and $\pi_k(0)=0,k\ge1$, the steady state probabilities of the system being empty can be obtained that $\pi_k=(1-\frac{\lambda}{\mu}){(\frac{\lambda}{\mu})}^k, k\ge0$. We specify this solution for $\frac{\lambda}{\mu}=\frac{1}{2}$ in Table \ref{tsolution}. Clearly from Table \ref{tsolution}, the steady state probability of $\pi_0$ is $0.5$, the steady state probability of $\pi_1$ is $0.25$ and the same is true of the rest of the states.

\begin{table}[htbp]
\caption{The solution for $\pi_k$ in an $M\backslash M \backslash 1$ Queue.}
\centering
\begin{tabular}{c||c|c|c|c|c}
\Xhline{1.2pt}
$k$&$0$&$1$&$2$&$3$&$4$\\
\hline
$\pi_k$&$0.5$&$0.25$&$0.125$&$0.0625$&$0.03125$\\
\hline
\hline
$k$&$5$&$6$&$7$&$8$&$9$\\
\hline
$\pi_k$&$0.0156$&$0.0078$&$0.0039$&$0.00195$&$0.00098$\\
\Xhline{1.2pt}
\end{tabular}
\label{tsolution}
\end{table}

\begin{figure}[htbp]
\centering
\includegraphics[width=8cm]{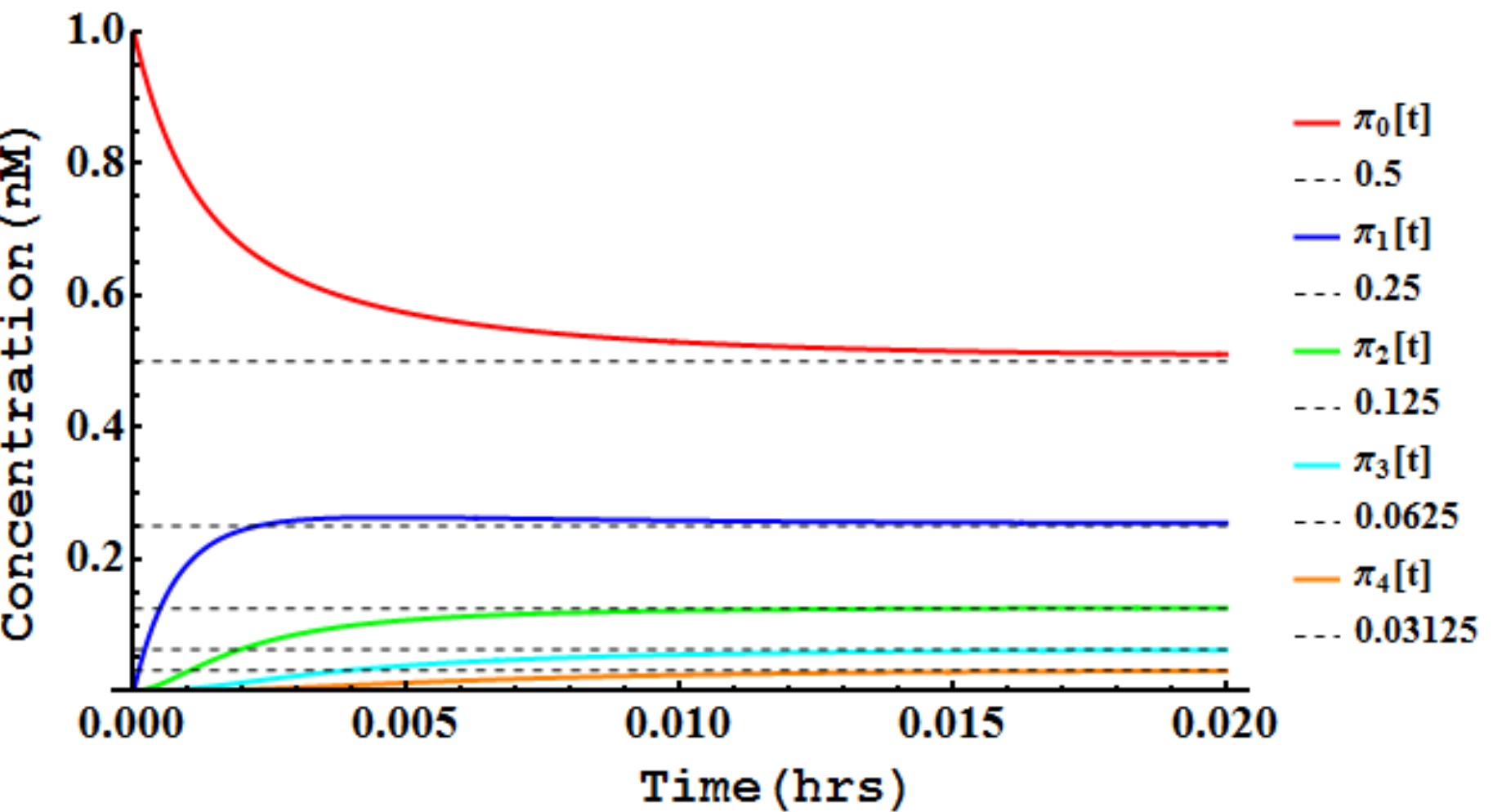}
\caption{ODE simulation result for a birth-death process.}\label{fig:birthdeathsim}
\end{figure}

The initial concentration of $\pi_0$ is unscaled $1$. The rate constants are unscaled $1$ and $2$. The scaled concentration and rates should be $10^{-9}$M, $0.1$/s and $0.2$/s. The simulation result is shown in Fig. \ref{fig:birthdeathsim}. Compared with Table \ref{tsolution}, the steady state probabilities are computed correctly by observing the final molecular concentrations. The maximal error gradually becomes less along with time as shown in Table \ref{terror}.

\begin{table}[htbp]
\caption{Error for $\pi_0$ in an $M\backslash M \backslash 1$ Queue.}
\centering
\begin{tabular}{c||c|c|c|c}
\Xhline{1.2pt}
t$(hrs)$&$0.02$&$0.04$&$0.06$&$0.08$\\
\hline
error$(\%)$&$2.163$&$1.598$&$1.588$&$1.587$\\
\Xhline{1.2pt}
\end{tabular}
\label{terror}
\end{table}

\begin{figure}[htbp]
\centering
\includegraphics[width=8cm]{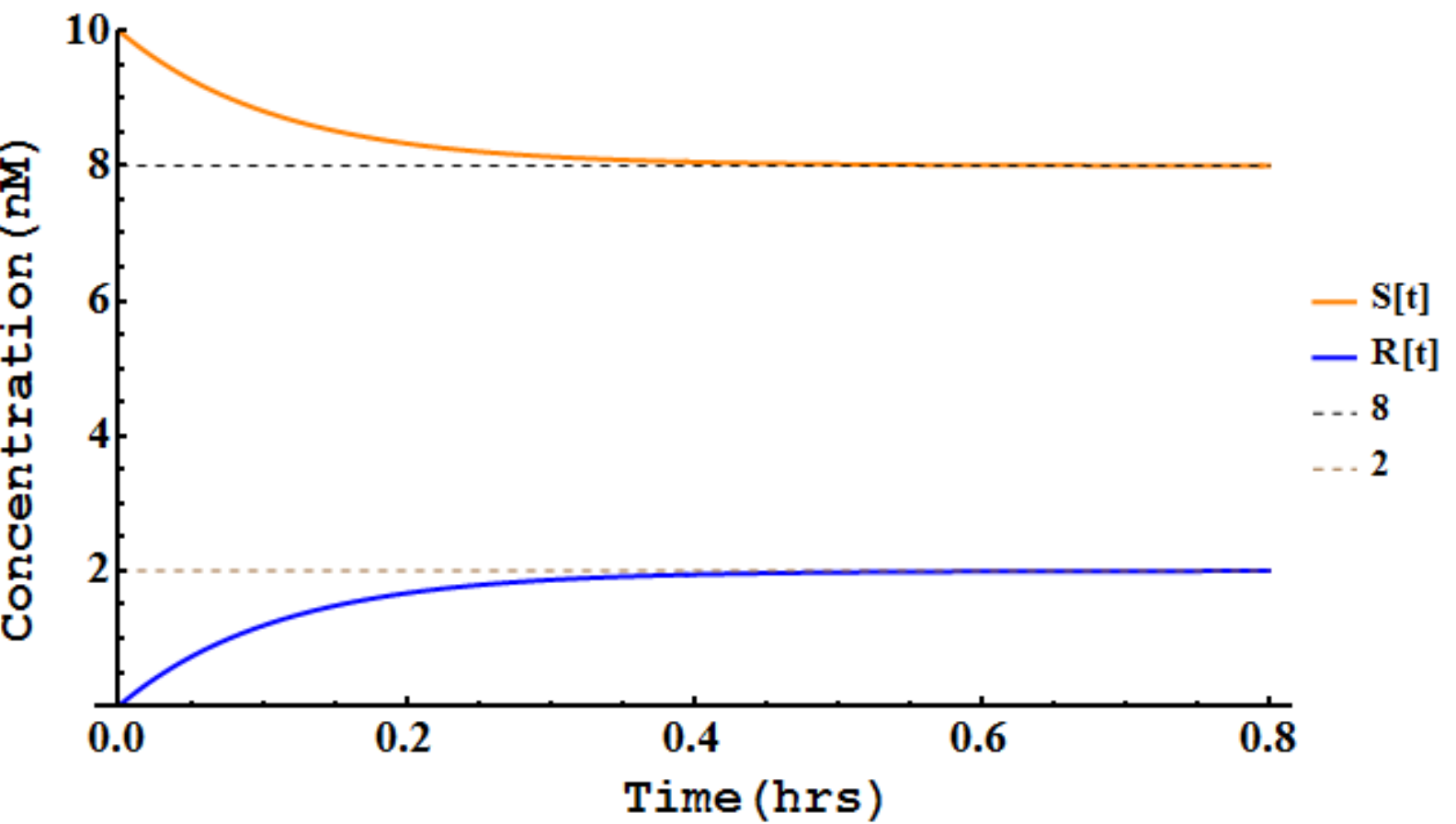}
\caption{ODE simulation result for weather prediction.}\label{fig:weatherode}
\end{figure}

\textbf{Weather Prediction:} If sunny is selected as the initial state, the initial concentration of $S$ is unscaled $1$. According to Fig. \ref{fig:weather}, the unscaled rate constants are $0.1,0.4,0.4,0.7$ as defined by transition probabilities. For simulations, the scaled concentration becomes $10^{-8}$M and the scaled rates are $0.05 \times 10^6$/M/s, $0.2 \times 10^6$/M/s, $0.2 \times 10^6$/M/s, $0.35 \times 10^6$/M/s. The simulation graph Fig. \ref{fig:weatherode} shows that the steady state distribution is $(0.8,0.2)$ as the equilibrium concentrations divided by the initial concentration are $0.8$ and $0.2$. The error is $0.45\%$ at hour $0.6$ and $0.07\%$ at hour $0.8$.

\subsection{Stochastic Simulation}
\begin{figure}[htbp]
\centering
\includegraphics[width=8cm]{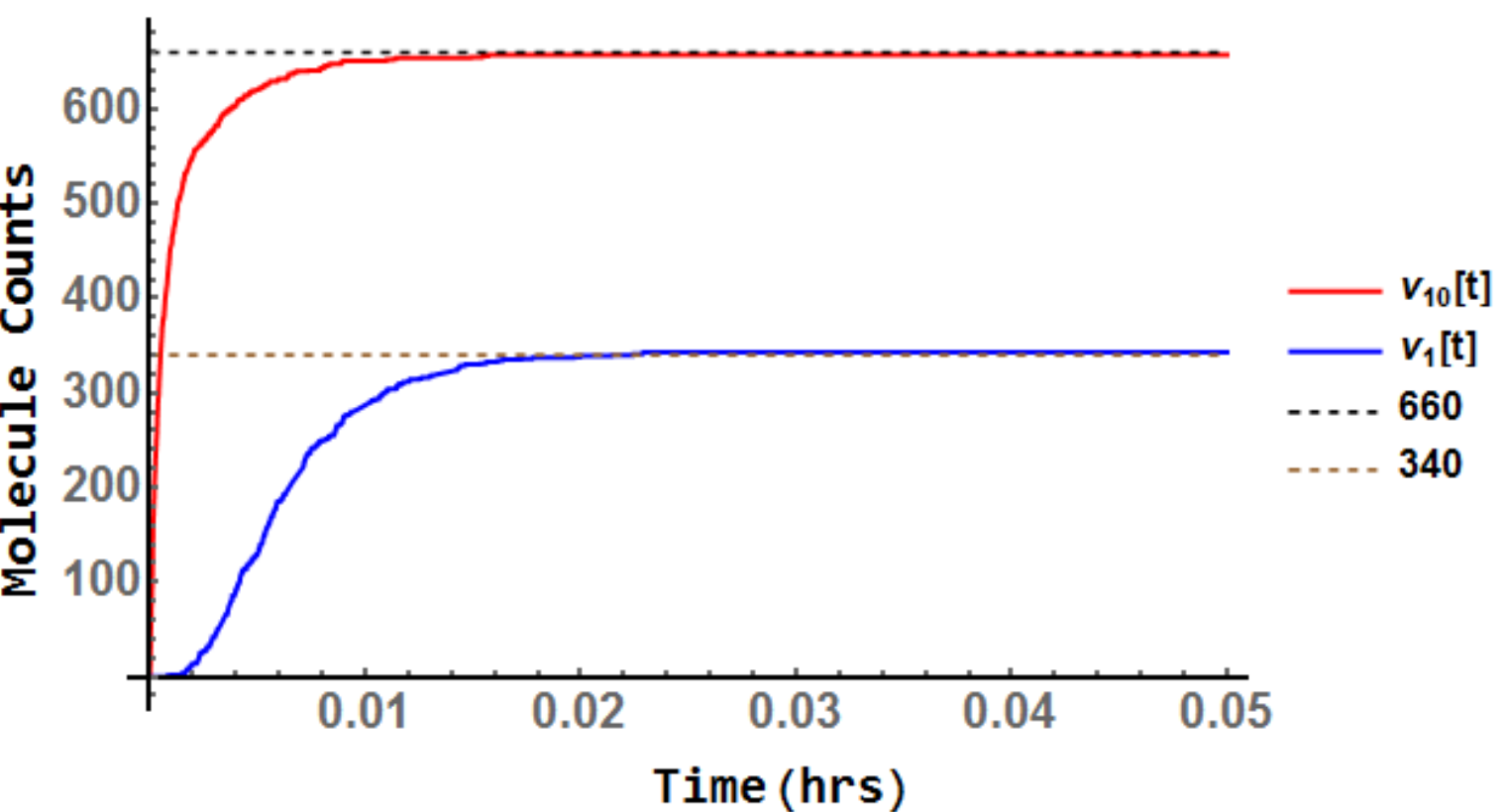}
\caption{Gillespie simulation result for gambler's ruin problem.}\label{fig:gamblergillespie}
\end{figure}

\begin{figure}[htbp]
\centering
\includegraphics[width=8cm]{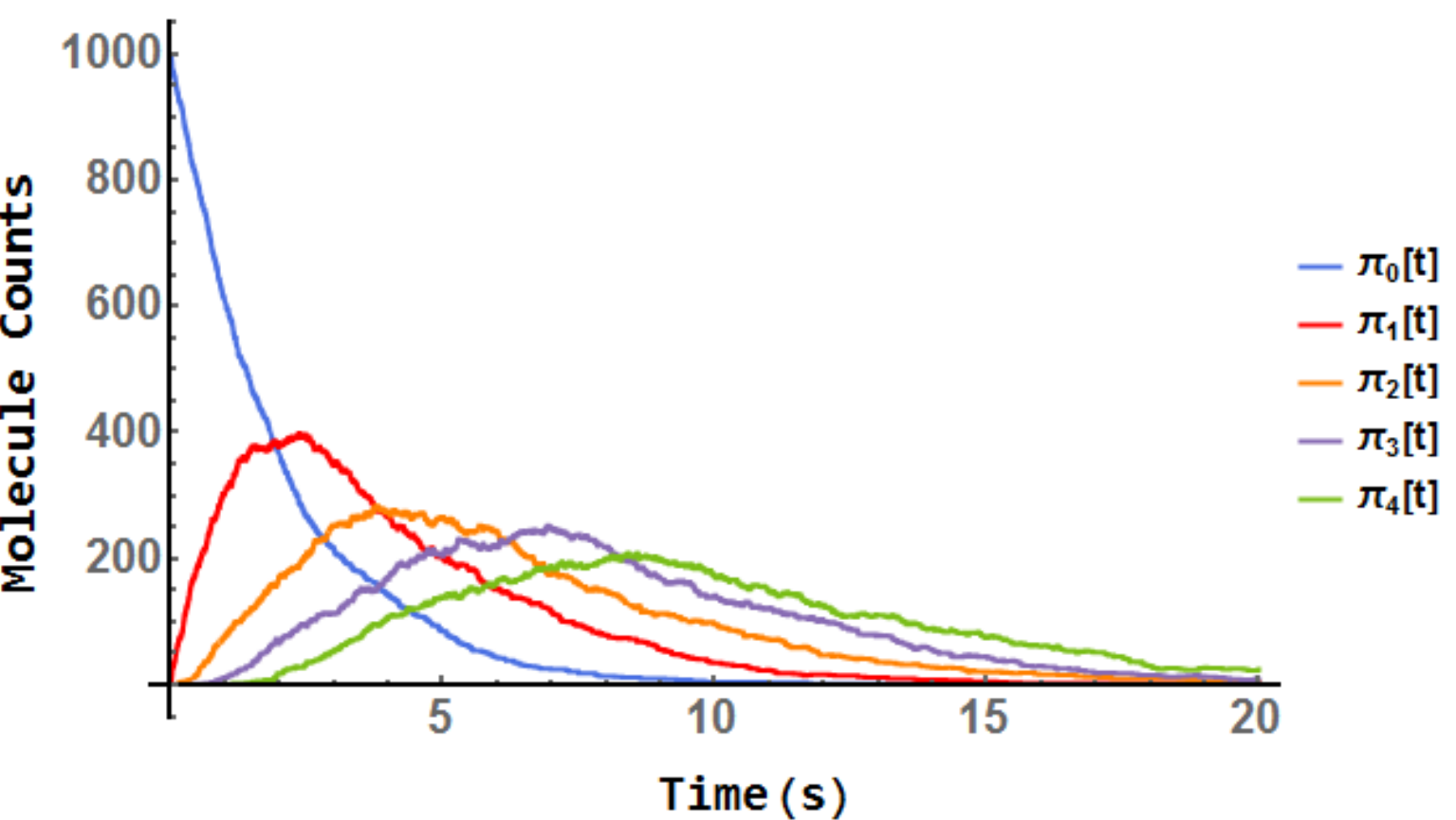}
\caption{Gillespie simulation result for a pure birth process.}\label{fig:purebirthgillespie}
\end{figure}

\begin{figure}[htbp]
\centering
\includegraphics[width=8cm]{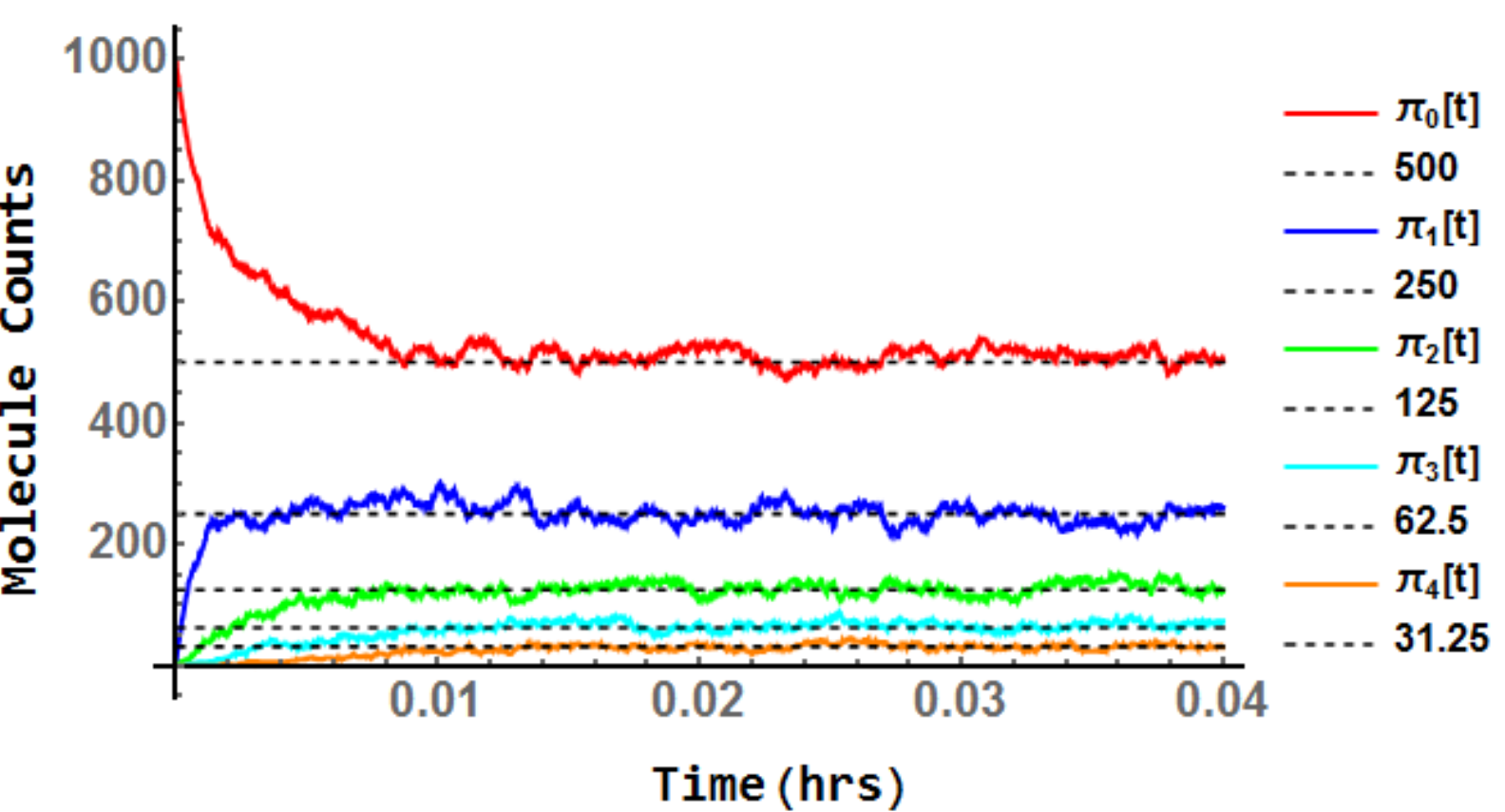}
\caption{Gillespie simulation result for a birth-death process.}\label{fig:birthdeathgillespie}
\end{figure}

\begin{figure}[htbp]
\centering
\includegraphics[width=8cm]{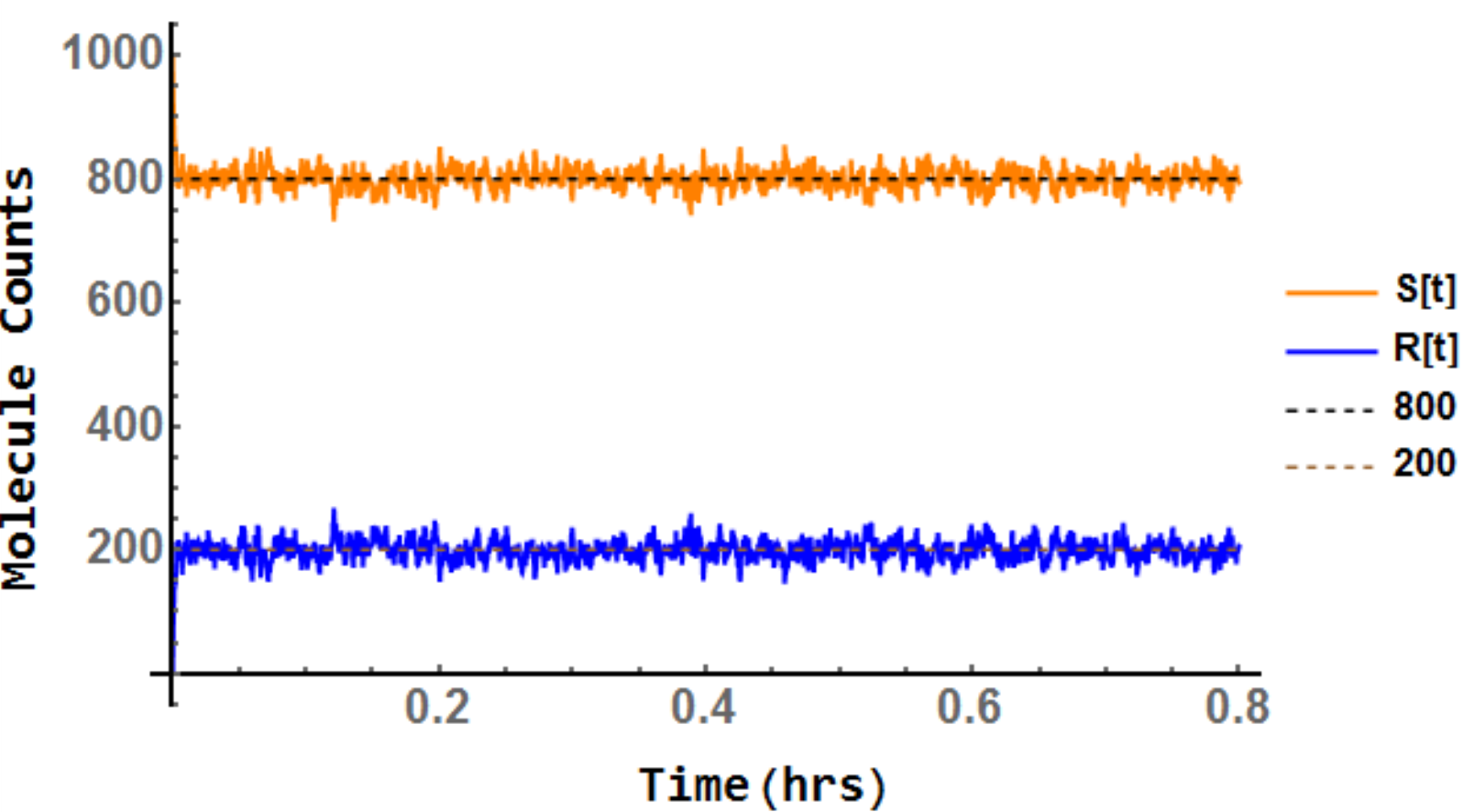}
\caption{Gillespie simulation result for weather prediction.}\label{fig:weathergillespie}
\end{figure}

Gillespie algorithm \cite{gillespie1976general} is used for stochastic simulation in this paper. Different from ODE simulation, gillespie simulation gives fluctuating curves as opposed to smooth ones. The result is indeterminate thus may deviate from the expected value. Nevertheless, the error can be reduced as the concentration increases. All initial numbers of molecules are selected to be $1000$ here for relatively accurate outputs. According to the simulation results in Fig. \ref{fig:gamblergillespie},\ref{fig:purebirthgillespie},\ref{fig:birthdeathgillespie},\ref{fig:weathergillespie}, concentrations vary above or below precise values in a limited range, effectively estimating the required outputs.

\section{DNA Implementation}
The methodology for an abstract set of molecular reactions is designed above. The engineered biochemical systems need to be mapped to specific DNA reactions to obtain real meaningfulness. In 2010, Soloveichik \cite{soloveichik2010dna} managed to contrive a DNA strategy for arbitrary hypothetical CRNs with satisfactory performance. In his method, each unimolecular reaction is compiled to two DNA strand displacement reactions and each bimolecular one is compiled to three. Buffering modules are additionally needed for bimolecular systems.

\subsection{Unimolecular Networks}
\begin{figure}[ht]
\centering
\includegraphics[width=16cm]{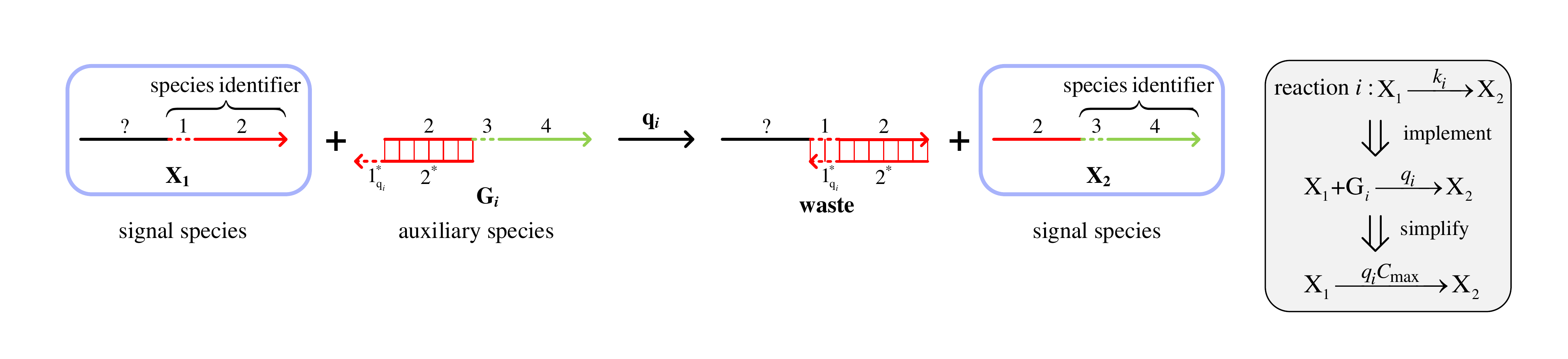}
\caption{DNA module for unimolecular networks.}\label{fig:dnauni}
\end{figure}

By carefully observing the network we design for first-order Markov chains, each reaction has one reactant and one product and the entire system contains only unimolecular reactions. Therefore, if Soloveichik's method is directly used here, it will be wasteful of DNA resources. Borrowing some clever ideas employed by Soloveichik, we devise a new DNA method for this typical unimolecular system as in Fig. \ref{fig:dnauni}. Each formal species is modeled by a kind of DNA strand named signal species, with the species identifier defined by one toehold and one domain. Each reaction is implemented by one DNA strand displacement reaction, with one signal species reacting with the auxiliary species $G_i$ to produce another signal species. The initial concentration of auxiliary species is $C_{max}$ and it is required that $max\{x_{X_j}(0)\}\ll C_{max}$. $q_i$ is the rate constant of the DNA reaction and it is controlled by the binding energy of domains $1_{q_i}^*$ and $1$, as $1_{q_i}^*$ is not a full complement of $1$. To ideally approximate the ODE kinetics, it should be satisfied that $q_iC_{max}=k_i$.

\begin{figure}[ht]
\centering
\includegraphics[width=16cm]{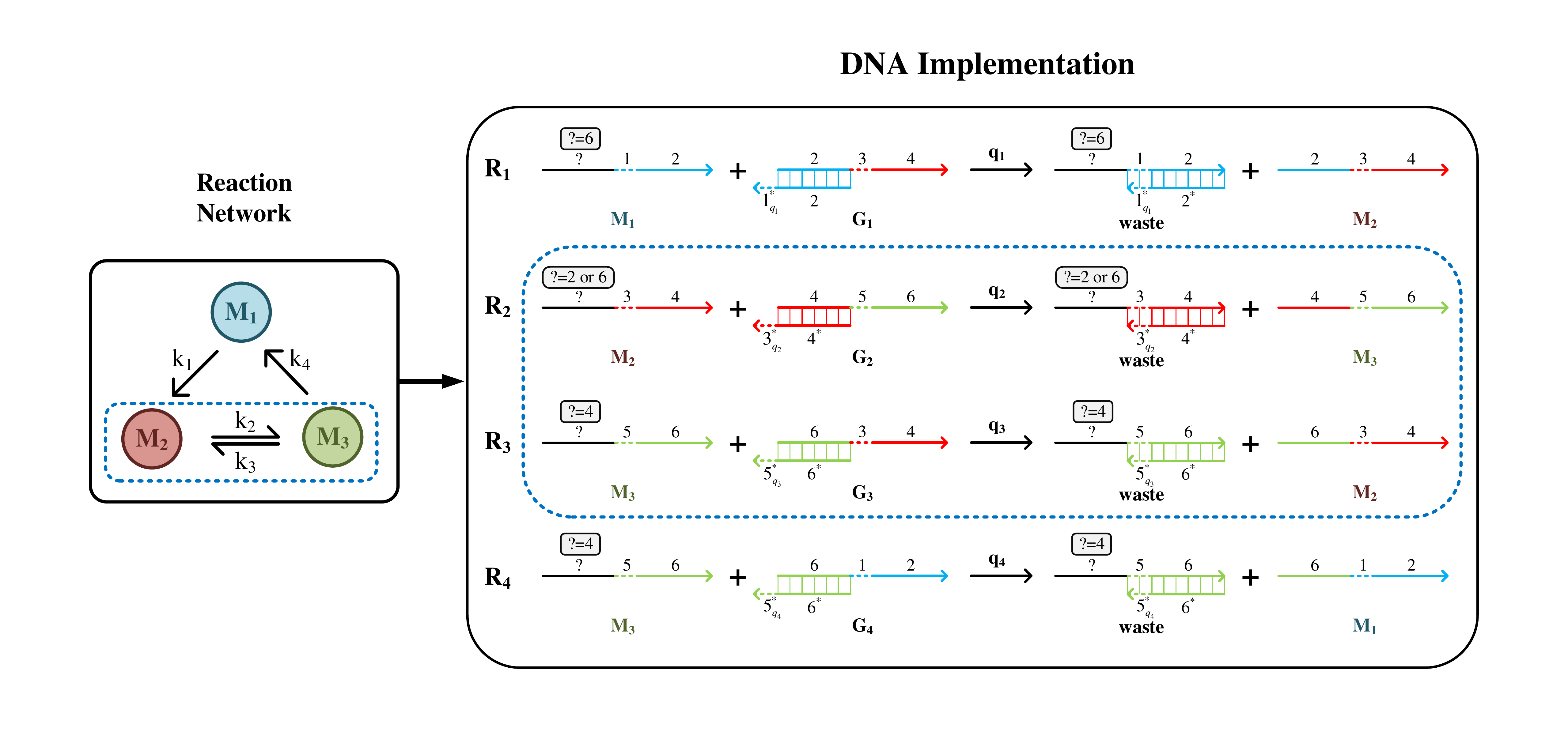}
\caption{DNA implementation of first-order Markov chains.}\label{fig:dnamodel}
\end{figure}

Utilizing a reaction network with three molecular species, a more specific mapping is shown in Fig. \ref{fig:dnamodel}. Given that different reactions may produce the same hypothetical species, the history domain ``$?$'' of each signal species is indeterminate. However, each signal species can be uniquely identified by the species identifier.

\begin{Rem}
By changing the length and sequence composition of a toehold domain $1_{q_i}^*$, which is not a full complement of $1$, the binding strength and in turn the rate constant can be varied. The rate constants can then be controlled over $6$ orders of magnitude \cite{yurke2003using,zhang2009control}. However, toeholds are short and have limited sequences. Although distributed over a wide range, not all exact rate constants can be achieved this way. To tackle this problem, concentrations of auxiliary species can be adjusted to fine-tune rate constants \cite{soloveichik2010dna}.
\end{Rem}

\begin{figure}[htbp]
\centering
\includegraphics[width=8cm]{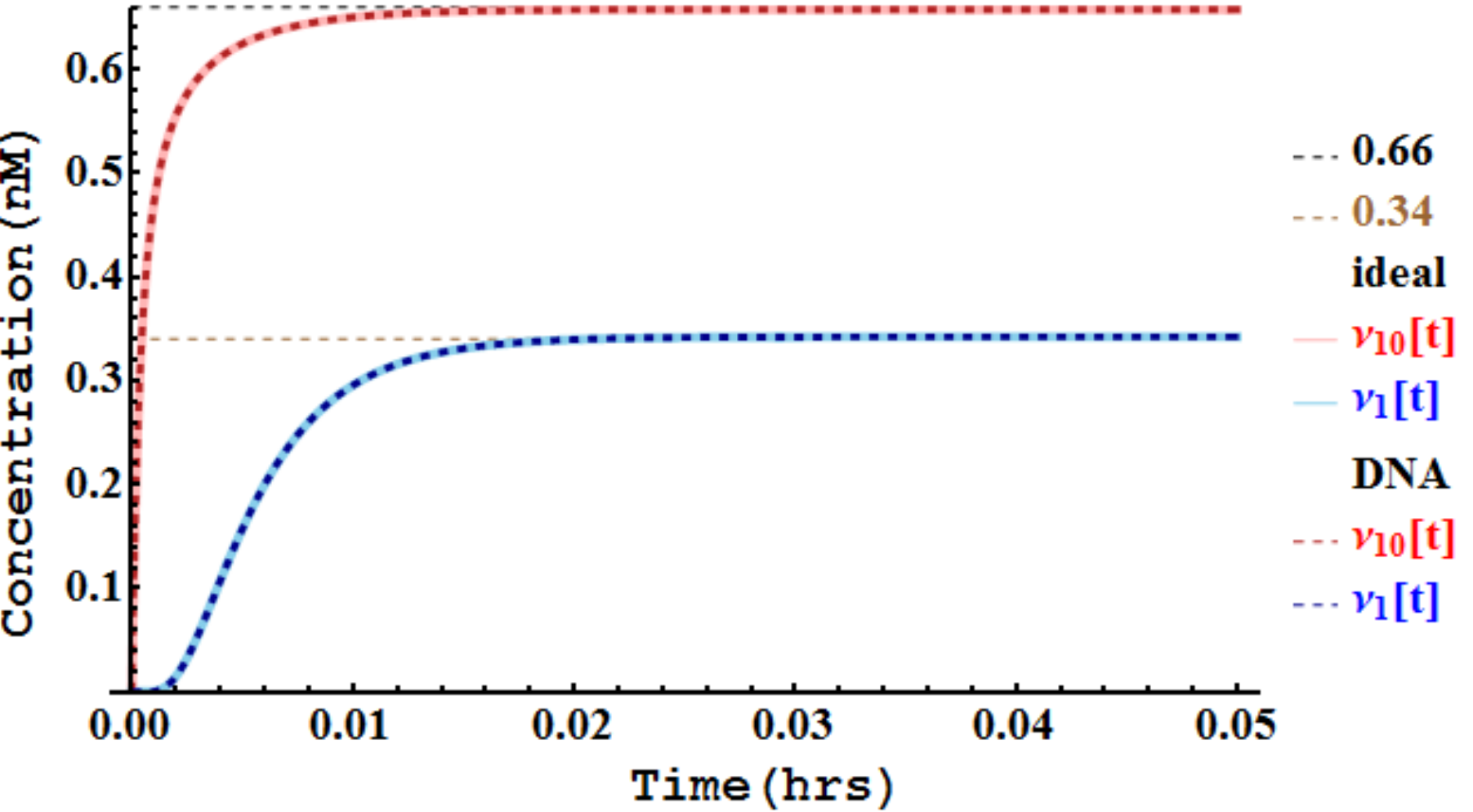}
\caption{DNA simulation result for gambler's ruin problem.}\label{fig:gamblercompare}
\end{figure}

\begin{figure}[htbp]
\centering
\includegraphics[width=8cm]{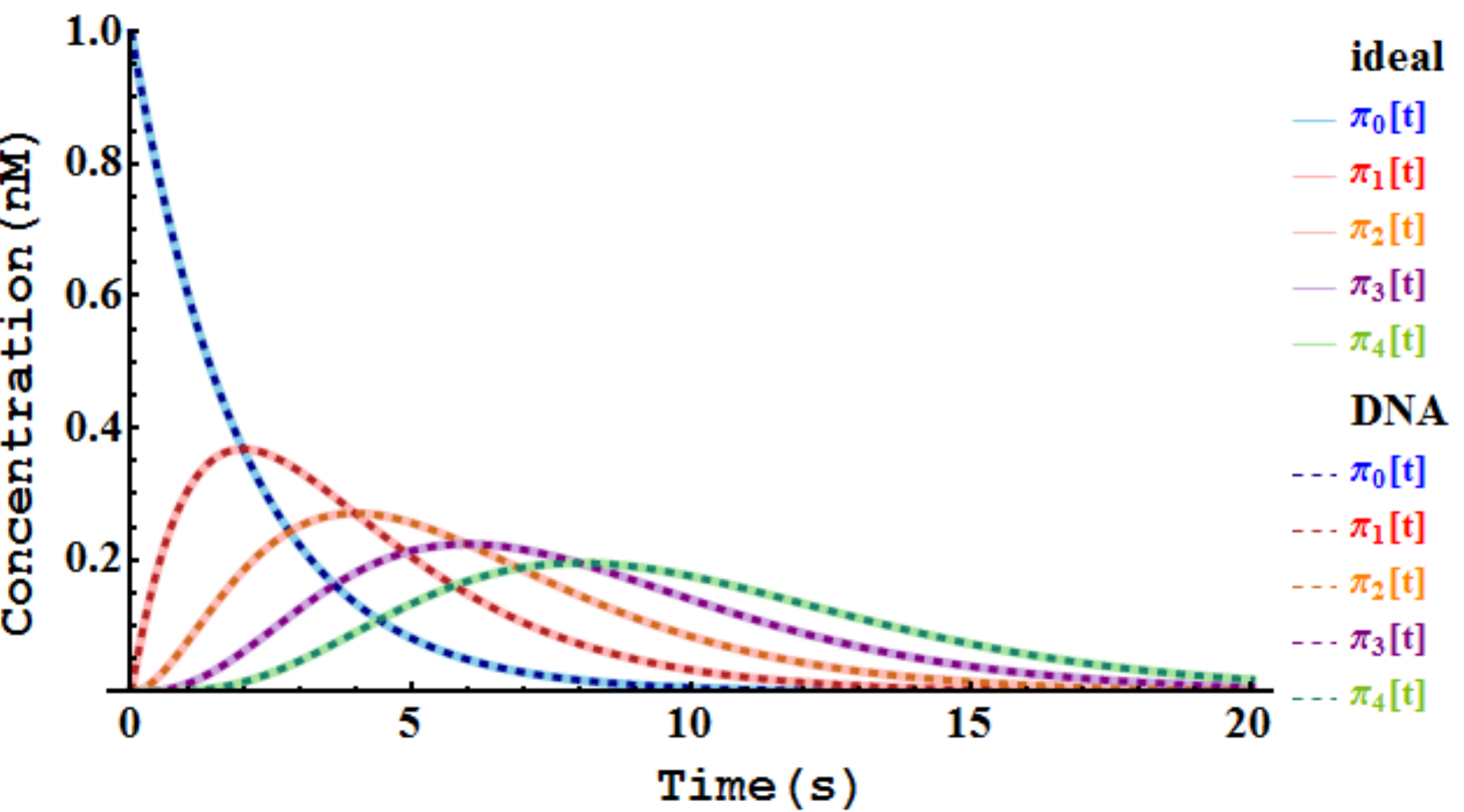}
\caption{DNA simulation result for a pure birth process.}\label{fig:purebirthcompare}
\end{figure}

\begin{figure}[htbp]
\centering
\includegraphics[width=8cm]{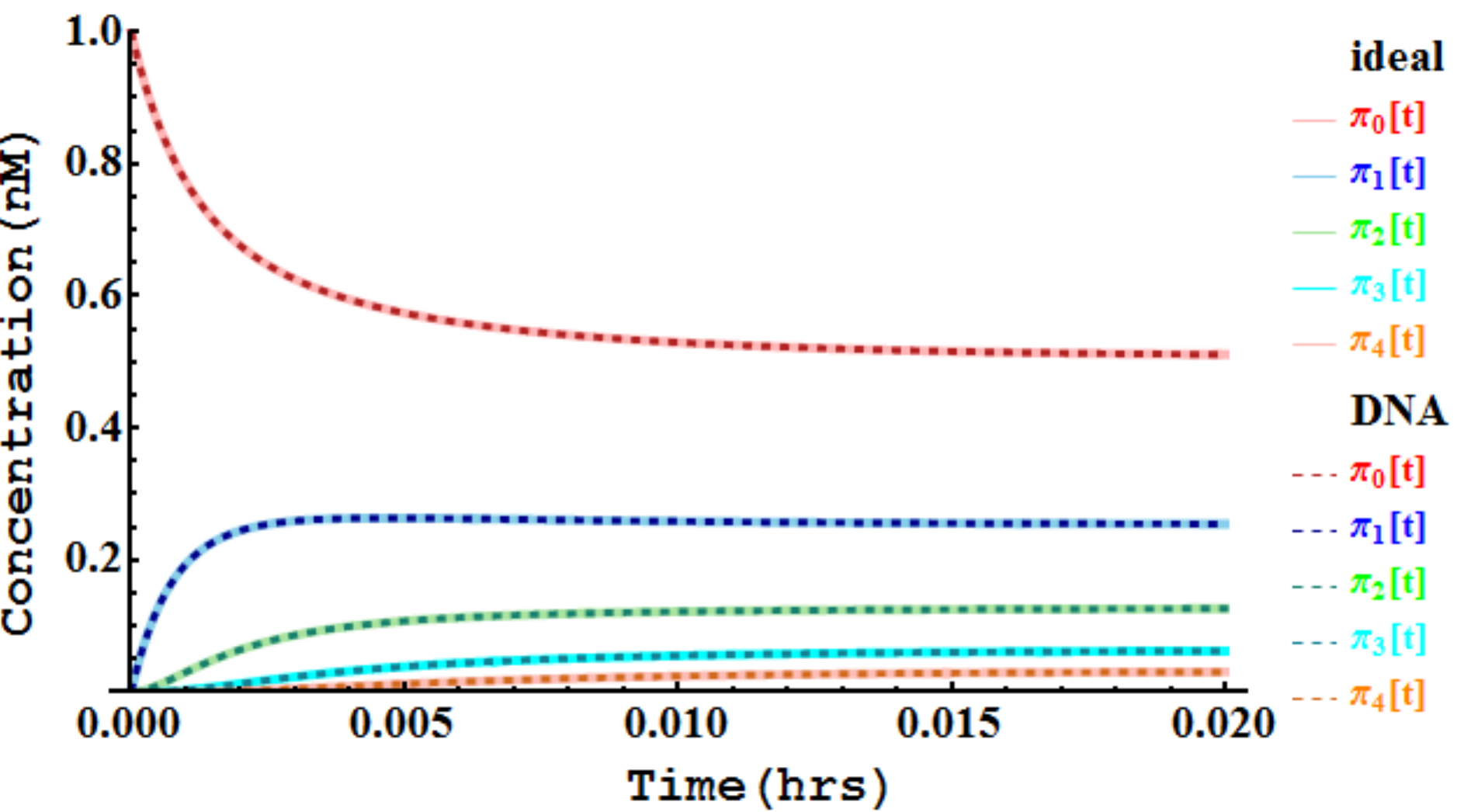}
\caption{DNA simulation result for a birth-death process.}\label{fig:birthdeathcompare}
\end{figure}

DNA simulations for the three examples with first-order chains are illustrated in Fig. \ref{fig:gamblercompare},\ref{fig:purebirthcompare},\ref{fig:birthdeathcompare}. Note that $C_{max}$ is set as $10^{-5}$M. DNA kinetics are drawn in dashed lines in contrast with ideal ODE kinetics. Compared to the ideal kinetic behaviors, those presented by DNAs are highly adequate.

\subsection{Bimolecular Networks}
\begin{figure}[htbp]
\centering
\includegraphics[width=8cm]{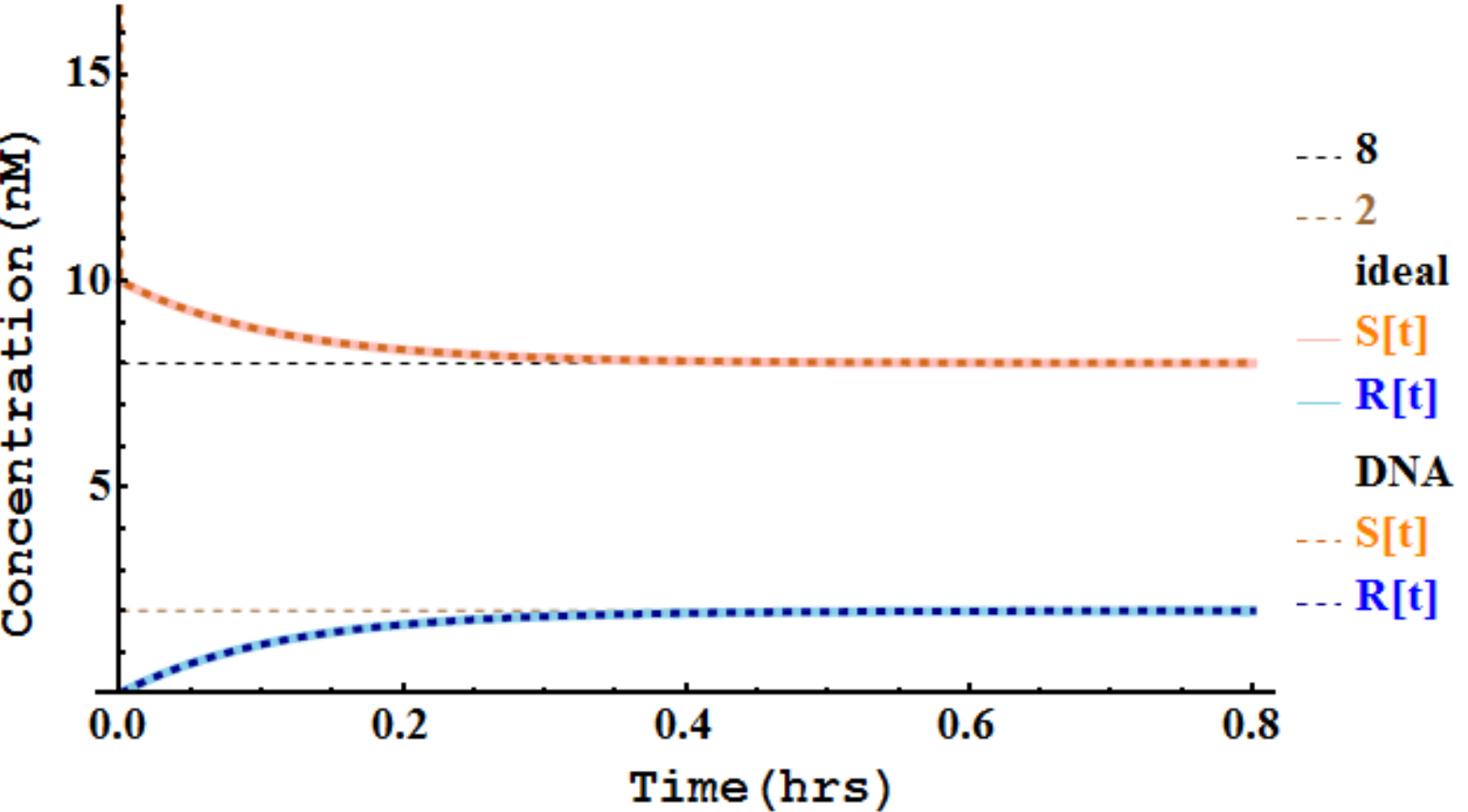}
\caption{DNA simulation result for weather prediction.}\label{fig:weathercompare}
\end{figure}

Reactions employed to realize second-order Markov chains have two reactants, thus the DNA approach proposed above is no longer effective. Technique proposed by Soloveichik \cite{soloveichik2010dna} is directly used here for simulation, where the species identifier is composed of one domain and two toeholds. The result of weather prediction is displayed in Fig. \ref{fig:weathercompare}. Notice that the initial concentration in the DNA system is $\frac{5}{3} \times 10^{-8}$M in that the buffering-scaling factor $\gamma^{-1}=\frac{5}{3}$.

\section{Conclusions}
In this paper, conjectural chemical reaction networks are shaped for computation of arbitrary time-homogeneous Markov chains, including DTMC, CTMC and second-order Markov chains. Not only steady state probabilities but also transient solutions are well synthesized. An original DNA method is proposed for implementing any unimolecular network with only one product in each reaction. Deterministic, stochastic and DNA simulations are provided to enhance correctness, validity and feasibility.



\bibliography{achemso-demo}


\end{document}